\documentclass{aa}  
\usepackage{graphicx}
\usepackage{subfig}
\usepackage{txfonts}
\usepackage[pdftex,usenames,dvipsnames]{color}
\usepackage{ulem}
\begin{document}

   \title{The s process in rotating low-mass AGB stars\thanks{This paper is dedicated to the celebration of the 100th birthday of prof.dr. Margaret Burbidge, in recognition of the outstanding contributions she has made to nuclear astrophysics.}}

   \subtitle{Nucleosynthesis calculations in models matching asteroseismic constraints}

   \author{J.W. den Hartogh
          \inst{1,2,3}
          \and
          R. Hirschi \inst{2,3,4}
          \and
          M. Lugaro \inst{1,5}
          \and 
          C.L. Doherty \inst{1,5}
          \and
          U. Battino \inst{6,3}
          \and
          F. Herwig \inst{7,8,3}
          \and 
          M. Pignatari \inst{9,1,3,8}
          \and
          P. Eggenberger \inst{10}
          }

   \institute{Konkoly Observatory, MTA CSFK, 1121, Budapest, Konkoly Thege Miklós út 15-17, Hungary\\
   \email{jacqueline.den.hartogh@csfk.mta.hu}
   \and 
   Astrophysics group, Lennard-Jones Laboratories, Keele University, ST5 5BG, UK
   \and
   NuGrid Collaboration, \url{http://www.NuGridstars.org}
   \and 
   Kavli Institute for the Physics and Mathematics of the Universe (WPI), University of Tokyo, 5-1-5 Kashiwanoha, 277-8583 Kashiwa, Japan
   \and
   Monash Centre for Astrophysics (MoCA), School of Physics and Astronomy, Monash University, Victoria 3800, Australia
   \and 
   School of Physics and Astronomy, University of Edinburgh, EH9 3FD, UK
   \and
   Department of Physics and Astronomy, University of Victoria, Victoria, BC V8P5C2, Canada
   \and 
   JINA-CEE, Michigan State University, East Lansing, MI, 48823, USA
   \and
   E. A. Milne Centre for Astrophysics, Department of Physics and Mathematics, University of Hull, HU6 7RX, UK
   \and
   Observatoire de Gen\`{e}ve, Universit\'{e} de Gen\`{e}ve, 51 Ch. des Maillettes, 1290 Sauverny, Suisse
             }

   \date{Received March 15 2019; accepted Augsut 20 2019}

  \abstract
   {}
   {We investigate the s-process during the AGB phase of stellar models whose cores are enforced to rotate at rates consistent with asteroseismology observations of their progenitors and successors.}
   {We calculated new 2M$_{\odot}$, Z=0.01 models, rotating at 0, 125, and 250 km/s at the start of main sequence. An artificial, additional viscosity was added to enhance the transport of angular momentum in order to reduce the core rotation rates to be in agreement with asteroseismology observations. We compared rotation rates of our models with observed rotation rates during the MS up to the end of core He burning, and the white dwarf phase.}
   {We present nucleosynthesis calculations for these rotating AGB models that were enforced to match the asteroseismic constraints on rotation rates of MS, RGB, He-burning, and WD stars. In particular, we calculated one model that matches the upper limit of observed rotation rates of core He-burning stars and we also included a model that rotates one order of magnitude faster than the upper limit of the observations. The s-process production in both of these models is comparable to that of non-rotating models.}
   {Slowing down the core rotation rate in stars to match the above mentioned asteroseismic constraints reduces the rotationally induced mixing processes to the point that they have no effect on the s-process nucleosynthesis. This result is independent of the initial rotation rate of the stellar evolution model. However, there are uncertainties remaining in the treatment of rotation in stellar evolution, which need to be reduced in order to confirm our conclusions, including the physical nature of our approach to reduce the core rotation rates of our models, and magnetic processes.}


\maketitle
%

\section{Introduction}
The asymptotic giant branch (AGB) is the last evolutionary phase of low- and intermediate-mass stars between approximately 0.8 to 10 M$_{\odot}$ before they evolve towards the white dwarf cooling track via the post-AGB and planetary nebulae phases \citep[see][]{2014PASA}. During the AGB phase the slow neutron capture process (s process) is activated \citep[see e.g.][]{B2FH,Kappeler2011}, and the stellar envelope becomes enriched with heavy elements before it is all lost to the interstellar medium by strong mass loss. \\
As a consequence, AGB stars contribute significantly to the galactic chemical evolution of the abundances of the elements beyond iron \citep[see e.g.][]{Travaglio2004,Bisterzo2017,Prantzos2018}. The main contributors are the low-mass AGB stars, which have an initial mass of 1.5-3 M$_{\odot}$.
A low-mass AGB star consists of a degenerate core, of mostly C and O, and a convective envelope \citep{falk_ARAA}. In between the core and the envelope, the H- and He-burning shells are active. The He-rich region between these two shells (the intershell) is where recurrent He flashes \citep[thermal pulses or TPs, first described by][]{1982IbenRenzini} take place. These TPs temporarily create a convective zone encompassing the whole intershell, which expands due to the flash. Due to the expansion, the H-free intershell cools and becomes convective, allowing for intershell material to be dredged up (third dredge-up or TDU) to the surface. The TDU also allows for H to be mixed into the intershell across the border of the convective H-rich region into the radiative He-rich region \citep{1998Gallino,2000herwig}. The $^{12}$C present in the intershell and the newly added H creates $^{13}$C via $^{12}$C(p,$\gamma$)$^{13}$N($\beta^{+} \nu$)$^{13}$C. The exact mixing process that allows for this creation of $^{13}$C is unknown, see \citet{Buntain2017} for a recent overview of the options, including convection and/or semiconvection \citep{1988Hollowell}, rotation \citep{Langer1999,2003ApJHerwig_rot,2004Siess}, diffusive convective boundary mixing \citep[CBM, as in][]{2000herwig,2013_fruity_rotation,Battino_2016}, gravity waves \citep{2003Denissenkov}, and magnetic buoyancy \citep{2014Nucci}. After a $^{13}$C-rich layer (or `$^{13}$C-pocket') has been created, neutrons are released via $^{13}$C($\alpha$,n)$^{16}$O in radiative regions during the interpulse period \citep{1995Straniero}. The neutrons are captured by iron-group elements, leading to the production of elements heavier than iron along the s-process path until the $^{13}$C is depleted and the next TP occurs. The TP ingests these elements, exposes them to a less significant neutron flux from the $^{22}$Ne neutron source, and mixes them up to the layers that will be mixed to the surface by the following TDU.\\
Most theoretical studies on AGB stars consider non-rotating models \citep[see][for reviews]{falk_ARAA,2006_Straniero_Gallino_Cristallo,2014PASA}. However, stars rotate. The effects of including rotation in stellar evolution calculations have been studied for almost a century, starting with \citet{Zeipel1924} and \citet{Eddington1925}. In short, rotation alters the stellar evolution in two ways: it deforms the stellar surface at high rotation rates, and it induces instabilities, which transport angular momentum and mix chemical elements \citep{2000ApJHeger,2000ARA&AMandM,ReviewMandM}. Pioneering work on rotating AGB stars was performed by \citet{Langer1999}, who found evidence that rotational mixing could be the mechanism needed to make a $^{13}$C-pocket. These rotating models were analysed in more detail by \citet{2003ApJHerwig_rot}, who found that the $^{13}$C-pocket created via rotationally induced mixing has not enough mass to achieve the overabundance in the envelope observed in AGB stars. Furthermore, \citet{2003ApJHerwig_rot} also found that in a 3 M$_{\odot}$ star of solar metallicity rotating with an initial rotational velocity of 250 km s$^{-1}$, the rotational mixing reduces the amount of neutrons available for the s-process \citep[as confirmed by][]{2004Siess}. The reduction is caused by the extra mixing of the neutron poison $^{14}$N \citep{2016Wallner} produced by $^{13}$C($p,\gamma)^{14}$N into the $^{13}$C-pocket during the long interpulse period. Consequently, their rotating AGB model did not lead to significant s-process production. \citet[the FRUITY models]{2013_fruity_rotation} presented the first set of yields for rotating AGB stars. As in \citet{2003ApJHerwig_rot,2004Siess}, these authors found that adding rotation leads to extra mixing within the $^{13}$C-pocket. However, they found that rotation does not necessarily eliminate the occurrence of the s-process and that it could produce a spread of s-process production patterns in AGB stars. The main difference between this study and \citet{2003ApJHerwig_rot} and \citet{2004Siess} is that the authors of \citet{2013_fruity_rotation} used lower initial rotation rates of 10, 30, 60, and 120 km s$^{-1}$ and also varied efficiency parameters of rotationally induced mixing, hence reducing the amount of extra mixing due to rotation, and its consequences for the s-process production.\\
While \citet{Langer1999}, \citet{2003ApJHerwig_rot}, and \citet{2004Siess} created the $^{13}$C-pocket via strong shear mixing at the bottom of the TDU, the $^{13}$C-pockets of the slower rotating models in this paper and in \citet{2013_fruity_rotation} are created by CBM.\\
In the meantime, new information has been gathered since 2012 on the internal rotation profile of low-mass stars resulting from asteroseismology studies of observations provided by the Kepler spacecraft \citep{2010borucki}. These studies provide values for the core rotation rates of low-mass stars that stellar evolution codes have been unable to match \citep[][]{2012eggenberger,Marques2013,2013tayar,2014cantiello}, confirming earlier findings of \citet[ e.g.][]{pin89,cha95,egg05,2008Suijs,Denissenkov2010}. In view of this mounting evidence that has accumulated over more than a decade, there is now consensus that a process of angular momentum transport is missing in the theory of rotating stellar evolution models. This missing process might also influence the s-process production in AGB stars. There is no widely accepted theory to explain this missing process, therefore constraints are required on its efficiency. \citet{2012eggenberger,Eggenberger_2017,Eggenberger2019}, and \citet[Paper I]{Jacqueline1} have characterised the efficiency of the mixing process of angular momentum transport by adding a constant additional artificial viscosity ($\nu_{\mathrm{add}}$) to the equation that describes the transport of angular momentum in a star. This $\nu_{\mathrm{add}}$ is calibrated to asteroseismically obtained rotation rates. In the present paper, we use a value of $\nu_{\rm{add}}$ close to that found in Paper I.\\
The aim of this paper is to compare the s-process production of non-rotating AGB models to rotating AGB models that have been enforced to match the asteroseismically measured core rotation rates. This is done by including a constant $\nu_{\mathrm{add}}$ to the transport of angular momentum. We stress that other approaches to reduce the core rotation rates have been studied, including an expression for the transport of angular momentum that has a dependence on the internal differential rotation \citep{2016Spada} and the combination of differential rotation in convective regions and magnetised winds \citep{2018tayar}. \\ In Sect.\thinspace\ref{sec:input_parameters} we introduce our methodology and the input parameters of our set of models. Sect.\thinspace\ref{sec:mainresults} is focussed on the s-process production. In Sect.\thinspace\ref{sec:concl} we present our final remarks. In appendix \ref{sec:app_sproc_noNUadd} we analyse and discuss the s-process production of the models that rotate too fast to match asteroseismically measured rotation rates.

\section{Physics of models}
\label{sec:input_parameters}
We use MESA revision 8845 \citep[see][for details on the code specifications]{2011_MESA_1,2013_MESA_2} as in Paper I. The input parameters match the settings of the NuGrid collaboration papers \citep[see][]{2016Pignatari,Battino_2016} for the calculations of a 2 M$_{\odot}$ star \citep[with initial abundances from][and scaled to Z=0.01]{1993grevesse}. In brief, we use the Schwarzschild criterion for the convective boundary placement, and the single exponentially decaying convective boundary mixing (CBM) within a diffusive mixing scheme as in \citet{1997herwig_overs}, except at the bottom of the convective envelope during TDUs and at the bottom of TPs, where we use the double exponentially decaying scheme as in \citet{Battino_2016}\footnote{This double exponential was determined for these two specific locations only, and should thus not be included during the whole interpulse as is done in \citet{Goriely2017}.}. In summary, this prescription of the diffusion coefficient consist of the treatment of \citet{1997herwig_overs}:
\begin{equation}
D_{\rm{CBM}}(\mathit{z})=D_{\rm{0}} \rm{exp}^{(-2\mathit{z}/\mathit{f}_{\rm{1}}\mathit{H}_{\rm{P0}})},
\end{equation}
where $D_{\mathrm{CBM}}$ is the diffusion coefficient for the CBM, and $D_{\mathrm{0}}$ is the diffusion coefficient at the border of the convective zone. $z$ is the distance from the Schwarzschild boundary into the radiative zone, and $f_1H_{\mathrm{P0}}$ the scale height of the CBM region. The double exponentially decaying prescription of \citet{Battino_2016}, based on the studies on gravity waves \citep{2003Denissenkov} and on 3D hydrodynamics \citep{2007herwig}, is activated when $D_{\rm{CBM}}$ falls below $D_{2}$ at distance $z>z_{2}$. The diffusion coefficient then becomes: 
\begin{equation}
D_{\rm{CBM}}=D_{2}\rm{exp}^{(-2(\mathit{z}-\mathit{z}_2)/\mathit{f}_{2}H_{\rm{P0}})},
\end{equation}
where all variables are as defined as above, $D_{\mathrm{2}}$ was calibrated by \citet{Battino_2016} to match diffusion coefficients found by earlier studies: 
the one of \citet{2007herwig} for the bottom of the TP, and the one found determined by \citet{2003Denissenkov} to account for mixing due to gravity waves below the convective envelope at the maximum Lagrangian downward extent of the TDU. \\
The evolution of the mass loss rate during the AGB phase is introduced using the mass loss formula of \citet{1995blocker}. We start with a mass-loss parameter of $\eta$=0.01, and when the TDUs have created a carbon-rich envelope, we increase the value to 0.04 as in \citet{2016Pignatari} and \citet{Battino_2016}. We increase the mass-loss parameter to 0.5 when convergence issues occur at the final stages of the AGB phase \citep[ad discussed by][]{Wood1986,Herwig1999,Sweigart1999,Lau2012}. Our models evolve from this restart onward into the white dwarf phase. \\
The network used in our MESA calculations includes 19 isotopes: neutrons, $^1$H, $^2$H, $^3$He, $^4$He, $^7$Li, $^7$Be, $^8$B, $^{12,13}$C, $^{13,14,15}$N, $^{16,17,18}$O, $^{19}$F, $^{22}$Ne, and $^{56}$Fe. A set of 27 reactions are then used to calculate the changes in the composition. These 27 reactions include all the pp-chain reactions and the CN- and NO-cycles, as well as the triple-$\alpha$ reaction and several $\alpha$-capture reactions: $^{12}$C($\alpha$,$\gamma)^{16}$O, $^{14}$N($\alpha$,$\gamma)^{18}$F(e$^{+}$,$\nu)^{18}$O, $^{18}$O($\alpha,\gamma)^{22}$Ne, $^{13}$C($\alpha$,n)$^{16}$O, and $^{19}$F($\alpha$,p)$^{22}$Ne. Together, the isotopes and reactions included are sufficient to track the energy generation of a low-mass AGB star with an initial mass of 2 M$_{\odot}$.

\subsection{Minor code modifications in MESA}
As in \citet{Battino_2016} we slightly altered the MESA source code: we exclude \textit{clipping} from our calculations, which means that small convective zones in our models can have a mixing length larger than their actual size \citep[see][for more details]{Battino_2016}. We also altered the implementation of the CBM to ensure the double exponential CBM only becomes active during the TPs and TDUs. Finally, we made a modification related to the implementation of opacities in MESA. As in \citet{2016Pignatari} and \citet{Battino_2016} we use the OPAL Type 2 opacities (Type 2 includes tables for enhanced mass fractions of C and O compared to solar scaled which are needed in the interiors of stellar interiors) throughout the evolution. To do this in revision 8845, we needed to adjust the MESA source code to cancel the blending of the two types of OPAL opacity tables as this blending created an opacity jump in the region of interest. 

\subsection{Rotation}
The implementation of rotation in MESA follows \citet{2000ApJHeger} and is summarised here. The calculation of transport of angular momentum ($j \propto \Omega r^2$) and the effect of rotation on the mixing of chemical elements is calculated via diffusion equations for the angular velocity and the mass fraction of chemical elements respectively:
\begin{align}
&\left( \frac{\partial \Omega}{\partial t}\right)_m=\frac{1}{j}\left( \frac{\partial}{\partial m}\right)_t \left[ (4\pi r^2 \rho)^2jD_{\rm{am}} \left( \frac{\partial \Omega}{\partial m}\right) \right]-\frac{2\Omega}{r}\left(\frac{\partial r}{\partial t}\right)_m\left( \frac{1}{2}\frac{d\rm{ln}\mathit{j}}{d\rm{ln}\mathit{r}}\right),  
\label{eq:AMtranport}\\
&\left(\frac{\partial X_{\rm{n}}}{\partial t} \right)=\left(\frac{\partial}{\partial \mathit{m}}\right)_{\rm{t}} \left[ (4\pi \rm{r}^2\rho)^2\mathit{D}_{\rm{mix}}\left ( \frac{\partial\rm{X_n}}{\partial m}\right)_{\rm{t}} \right]+\left(\frac{\rm{dX_n}}{\rm{d}t}\right)_{\rm{nuc}}.
\label{eq:Mixing_Xi}
\end{align}
where $\Omega$ is the angular velocity, $j$ the specific angular momentum, and the total diffusion coefficient $D_{\rm{am}}$ takes into account all processes that transport angular momentum, which are given below. In the Eq. \ref{eq:Mixing_Xi}, $X_{\rm{n}}$ is the mass fraction of species n, and $D_{\rm{mix}}$ the sum of all processes that mix the chemical elements. The final term in Eq. \ref{eq:Mixing_Xi} accounts for the changes in composition due to nuclear reactions. The different diffusion coefficients $D$ are defined as follows:
\begin{align}
    &D_{\rm{am}}=D_{\rm{conv}}+D_{\rm{rot}}+\nu_,
    \label{eq:Dam}\\
    &D_{\rm{mix}}=D_{\rm{conv}}+\mathit{f}_c D_{\rm{rot}},\:\mbox{and }
    \label{eq:Dmix}\\
    &D_{\rm{rot}}=D_{\rm{ES}}+D_{\rm{SSI}}+D_{\rm{DSI}}+D_{\rm{SH}}+D_{\rm{GSF}}.
\end{align}
The individual terms correspond to convection (conv), the Eddington-Sweet (ES) circulation \citep[also known as meridional circulation:][]{Kippenhahn1974}, dynamical and secular shear instabilities (DSI and SSI, respectively) \citep{Zahn1974,endal_sofia1978}, the Solberg-H{\o}iland (SH) instability \citep{Wasiutynski1946}, and the Goldreich-Schubert-Fricke (GSF) instability \citep{Goldreich1967,Fricke1968}. The additional viscosity $\nu_{\rm{add}}$ is added to $D_{\rm{am}}$ only, and not to the mixing of chemical elements. This implementation is identical to the implementation of the $\nu_{\rm{add}}$ in Eggenberger et al. (2012,2017,2019) and Paper I. Arguments behind the exclusion of several of these instabilities in our calculations are given in the following subsection.

\subsubsection{Settings of rotationally induced instabilities}
\label{sec:rot_sett}
There are two free parameters in the implementation of rotation in MESA: the first is $f_c$ in Eq. \ref{eq:Dmix}, which allows the user to vary the contribution of the rotationally induces instabilities to the mixing of chemical elements. The second is $f_{\mu}$, which is added in front of the molecular weight gradient $\nabla_{\mu}$ that appears in the determination of several of the instabilities in $D_{\rm{rot}}$ \citep[see][for the full definitions]{2000ApJHeger}. The parameter $f_{\mu}$ determines the dependence of the individual instability on the molecular weight gradient. Both $f$-parameters are introduced to compensate for various simplifications in the derivation of the instabilities, and are limited to values between 0 and 1. Our choice for these factors are explained in the following subsection.\\
The values of $f_{c}$ and $f_{\mu}$ are set to 1/30 and 0.05 respectively by \citet{2000ApJHeger}, based on theoretical work by \citet{1992chaboyer} and the calibration by \citet{2000ApJHeger} of the surface enrichment of nitrogen in massive stars at the end of the main sequence. The dependence on these two parameters of the s-process production in low-mass AGB stars has been investigated by \citet{2004Siess} and \citet{2013_fruity_rotation}. \citet{2004Siess} varied $f_{\mu}$ between 0$-$0.05 and found that $f_{\mu}$=0 leads to no s-process production even for very slow rotators, while slow rotators with $f_{\mu}$=0.05 results in s-process production. \citet{2013_fruity_rotation} found that varying $f_{\mu}$ between 0.05$-$1 and $f_c$ between 0.04$-$1 results in variation in s-process production similar to the spread of s-process production obtained by changing the initial rotation rate between 10 and 120 km s$^{-1}$. We do not repeat these parameter studies and use the same values as \citet{2000ApJHeger}. This is because we now know from asteroseismology observations that a process of angular momentum transport is missing from the implementation of rotation in stellar evolutionary codes, which currently eliminates possibility of calibration. We further discuss these $f$ parameters in Sect.\thinspace\ref{sec:concl}. We also note that we do not include any type of smoothing of the diffusion profiles of the instabilities in our calculations.\\ 
Our standard rotating models include only the ES circulation and the SSI. We exclude all dynamical instabilities (DSI and SH) as these instabilities do not transport angular momentum (see Appendix B in Paper I) or participate in the mixing of chemical elements. We discuss this point in more detail in Appendix \ref{sec:app_sproc_noNUadd}. The exclusion of the GSF instability is based on \citet{2010Hirschi} and \citet{2016caleo}. The first paper shows that the GSF instability is not responsible for the low rotation rates of pulsars, while the second focusses on the Sun and 1.3 M$_{\odot}$ RGB stars and shows that the GSF instability is unlikely to be activated in those stars. Both papers show that viscosity, assumed to be negligible in the original derivation of the instability \citep{1970James,1971James}, either turbulent as in \citet{2010Hirschi} or molecular and radiative as in \citet{2016caleo}, suppresses the GSF instability. \citet{2010Hirschi} shows that for several evolutionary phases of a 20 M$_{\odot}$ star, the GSF instability is always weaker than the dynamical shear. The implementation of the GSF instability in MESA currently follows \citet{2000ApJHeger} and does not include the stabilising effect of the viscosity. Therefore, we exclude GSF from our simulations.

\subsection{The post-processing tool MPPNP and the reaction rate network}
\label{sec:mppnp}
We use the NuGrid multi-zone post-processing tool MPPNP \citep[see][for details on the code and the reaction rate network]{Bennett2012,2016Pignatari}. MPPNP uses temperature, density and diffusion output of MESA to calculate the nucleosynthesis during the whole stellar evolution. As the sum of all diffusion processes is included in the MESA output, we can use MPPNP also for rotating models. The network and corresponding reaction rates used are the same as in \citet{Battino_2016}.\\

\begin{figure}[t]
\centering
\includegraphics[width=\linewidth]{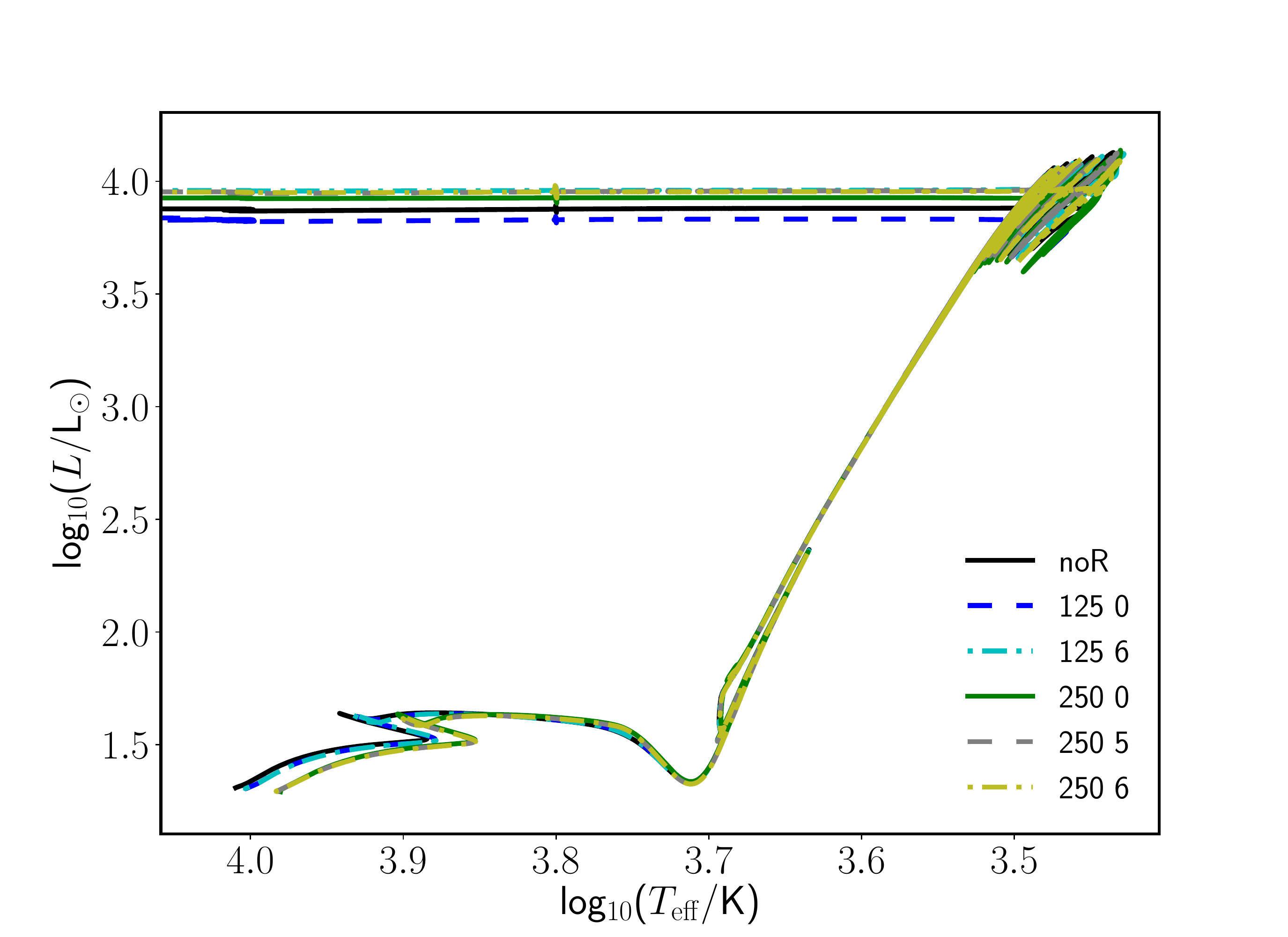}
\caption{The HRDs of all our non-rotating and rotating models, see text for discussion.}
\label{fig:hrd_zoomin}
\end{figure}

\begin{table}
\centering
\caption{Model properties. Names of the models are a combination of initial rotation rate (first number) and the order of magnitude of $\nu_{\rm{add}}$ (second number). The H-free core mass and the core rotation rate are given at the time the first TP (TP1) occurs. The white dwarf mass ($M_{\rm{DAV}}$) is taken when the star proceeds through the DAV phase (see text for details) on the white dwarf cooling track.}
\begin{tabular}{c|ll|l|ll}
Run&$v_{\rm{rot,i}}$& $\nu_{\rm{add}}$ &  $\Omega_{\rm{c,TP1}}$ & $M_{\rm{DAV}}$ & $\Omega_{\rm{DAV}}$\\
 &   \small{km s$^{-1}$} & \small{cm$^2$ s$^{-1}$} &  \small{2$\pi$nHz} & \small{M$_{\odot}$} & \small{day} \\
\hline 
noR    & - & - & & 0.62 &\\ 
\hline 
125 0 & 125 & 0      & 5.0$\times$10$^6$ & 0.62 & 1.5$\times$10$^{-3}$\\  
125 6 & 125 & 10$^6$ & 3.5$\times$10$^3$ & 0.62 & 1.3\\ 
\hline 
250 0 & 250 & 0      & 6.7$\times$10$^6$ & 0.62 & 1.1$\times$10$^{-3}$\\  
250 5 & 250 & 10$^5$ & 5.0$\times$10$^4$ & 0.62 & 0.11\\
250 6 & 250 & 10$^6$ & 5.2$\times$10$^3$ & 0.61 & 1.0\\
\end{tabular} 
\label{tab:mass}
\end{table}

\subsection{Set of models}
Our set of models is listed in Table \ref{tab:mass}. We calculated 2 M$_{\odot}$ models at metallicity Z=0.01. We chose two initial rotation rates set at the ZAMS: 125 and 250 km s$^{-1}$ corresponding to a v/v${_\mathrm{crit}}$ of 0.27 and 0.57 respectively. These initial values match the range found for very young B stars (log g$_{\rm{polar}}$ > 4.15) by \citet{2010Huang} and are similar to those used in previous publications of rotating AGB stars: \citet{Langer1999,2003ApJHerwig_rot,2004Siess} used 250 km/s for their 3 M$_{\odot}$ model, while \citet{2013_fruity_rotation} used up to 120 km s$^{-1}$ for their 2 M$_{\odot}$ star. \\
The value of 10$^6$ cm$^2$ s$^{-1}$ for $\nu_{\rm{add}}$ is chosen to reach the observed core rotation rates, see Fig.\thinspace\ref{fig:omega4}. In all models, $\nu_{\rm{add}}$=0 from the end of the core He burning phase onward. These settings follow the results of Paper I, except that the values used for $\nu_{\mathrm{add}}$ are lower than in Paper I. This difference is caused by the different aims of the papers: in Paper I we focussed on the observations of a small data set of core He burning stars \citep{Deheuvels2015}, while in this study we are interested in obtaining a model that can serve as an upper limit of all observed core rotation rates. We also include models with $\nu_{\rm{add}}$=0 for both initial rotation rates.\\
Fig.\thinspace\ref{fig:hrd_zoomin} shows the HRDs of the models listed in Table \ref{tab:mass} up to the post-AGB phase. Rotating models are located to the right of the non-rotating model on the ZAMS due to the centrifugal force expanding the star and producing a cooler surface. The core masses of the rotating models without $\nu_{\rm{add}}$ at the end of the main sequence are slightly larger than those of the models including $\nu_{\rm{add}}$ and of the non-rotating models because of the mixing of extra fuel into the core during the main-sequence. As a result of the larger core mass, the next core burning phase is shorter and therefore the core masses after the core He burning phase are comparable. Small variations in core masses occur after the AGB phase due to differences in the number of TPs and thus core growth during the AGB phase. This mass difference, see Table \ref{tab:mass} is visible as difference in luminosity in the post-AGB tracks in Fig.\thinspace\ref{fig:hrd_zoomin}.\\

\subsection{Rotational evolution}
\begin{figure}
    \centering
    \includegraphics[width=\linewidth]{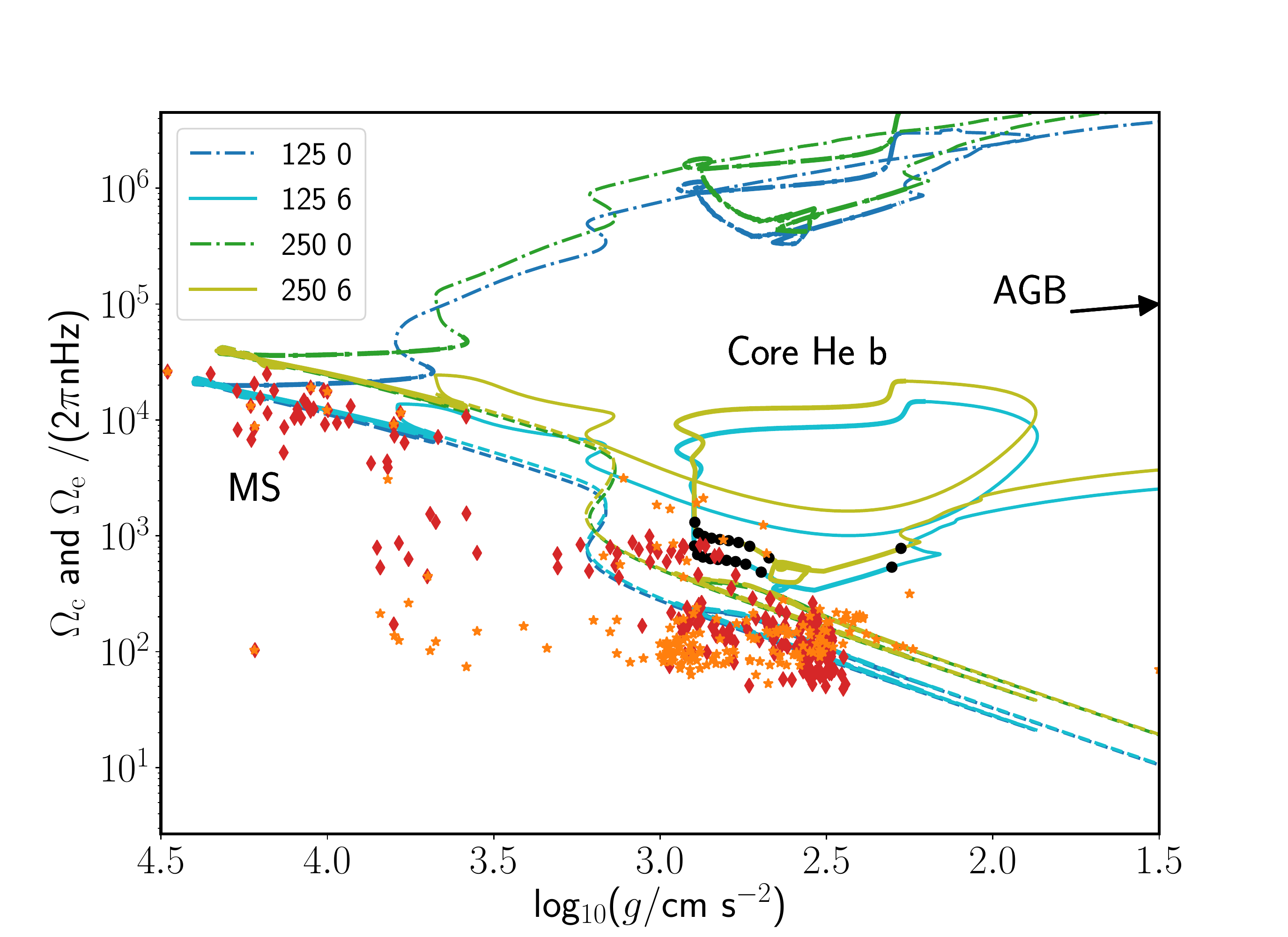}
    \caption{Evolution of core ($\Omega_c$) and surface rotation ($\Omega_s$) rates. Four of the models listed in Table \ref{tab:mass} are shown here, and compared to asteroseismically obtained rotation rates. The observational data points are the core (red diamonds) and the surface (orange stars) rotation rates, taken from \citet{2012Mosser}, \citet{2012deheuvels}, \citet{Deheuvels2014}, \citet{Deheuvels2015}, \citet{Ceillier2017}, and the compilation of observed main-sequence stars from 12 other papers presented in \citet{Aerts2017}. From these observational studies, we only select single stars in the mass range 1.4$-$3.0 M$_{\odot}$. Typical error bars of these observations are of the order of the symbol size used. The solid and dot-dashed show the core rotation rates of the models with and without the additional viscosity respectively. The dashed lines show the envelope rotation rates of the models. The thick line segments correspond to the core burning phases, and the thin segments to the shell burning phases. The black dots indicate the time spend in the core He burning phase by the models with $\nu_{\mathrm{add}}\neq$ 0, each spaced by 10\% of the total duration starting at the 10\% mark and ending with the 100\% mark (the dots located on the most left and right, respectively). These dots show that these models spend most of their time during this evolutionary phase close to the observed rotation rates. }
    \label{fig:omega4}
\end{figure}

Figure \ref{fig:omega4} shows four models from Table \ref{tab:mass}: the two rotating models without $\nu_{\mathrm{add}}$ and the two rotating models with $\nu_{\mathrm{add}}$=10$^6$ cm$^2$ s$^{-1}$. The different trends visible in the models with and without $\nu_{\rm{add}}$ are explained in detail in Paper I. In short, by adding $\nu_{\rm{add}}$, coupling is provided between the core and envelope that allows for transport of angular momentum from the core to the envelope, even during the evolutionary phases where the core is contracting. As a result, the core rotation rate shows a steady decrease during the evolution, instead of an increase as in the standard rotating model without $\nu_{\rm{add}}$.\\
From the four models shown in this figure, those with $\nu_{\rm{add}}$ = 0 only match the observations at the start of the main sequence, while those with $\nu_{\rm{add}}$=10$^6$ cm$^2$ s$^{-1}$ represent rough upper limits of the observed core rotation rates. During the core He burning phase the comparison between these models and the observations is especially important. We therefore added markers (black dots) to the two models, indicating every 10\% of the total duration of the core He burning phase. These dots show that from 10\% to 80\% of the total duration of the core He burning phase, the models are in the same location as the observed rotation rates in this figure.\\
The core rotation rates at the first TP are given in Table \ref{tab:mass}. These rates show a difference of three orders of magnitude between the models with and without $\nu_{\rm{add}}$. We also calculated a model with a low initial rotation rate of 10 km/s, which has a core rotation rate of $\Omega/2\pi$ = 7.88$\times$10$^5$ nHz at the first TP. This is still two orders of magnitude higher than the rotation rates of the models matching the observed rotation rates, showing that simply reducing the initial rotation rate cannot match the observed rotation rates. Another method to reduce the core rotation rate, for instance $\nu_{\mathrm{add}}$, is needed.\\
We also calculated the `250 5' model. The core rotation rate at the first TP is an order of magnitude higher than the `250 6' model. The core rotation rate during core He burning of this model is at least an order of magnitude higher than all observed core rotation rates for this evolutionary phase. At the first TP, the core rotation rate is an order of magnitude larger than the `250 6' model. Therefore, s-process production of this model can be considered a conservative prediction for the s-process production of stars rotating at rates matching the asteroseismically measured rotation rates.
In Table \ref{tab:mass} we also show our white dwarf rotation rates. Most of the white dwarfs for which rotation rates are known are DAVs, which are pulsating H-rich white dwarfs. They have a $T_{\rm{eff}}$ between 10600$-$12600 K, because the H on their surface has to be partially ionised for the pulsations to take place. Our presented rotation rates are taken within the DAV temperature range\footnote{The `250 5' model undergoes a very late thermal pulse (VLTP) whilst on the WD cooling track, before the DAV temperature range is reached. As this model runs into convergence issues before returning to the white dwarf track, we have taken the rotation rate just before the very late thermal pulse.}. As in Fig.\thinspace\ref{fig:omega4}, the models including $\nu_{\rm{add}}$ = 10$^6$ cm$^2$s$^{-1}$ match the observed white dwarfs rotation rates from \citet{2015kawaler} and \citet{Hermes2017}, while the `250 5' model is an order of magnitude too low. The models without $\nu_{\rm{add}}$ are far from the observed values (confirming the results of \cite{2008Suijs} and \cite{2014cantiello}). As mentioned previously, we remove $\nu_{\rm{add}}$ after the end of the core He burning phase, therefore conserving angular momentum within the core from this point onward. As in \citet{2014cantiello,Aerts2018}, this approach allows to match the observed rotation rates during both the core He burning phase and the white dwarf cooling track.\\

\section{s-process production in models matching asteroseismically measured rotation rates}
\label{sec:mainresults}

\begin{figure*}[ht]%
    \centering 
    \subfloat{{\includegraphics[width=6.15cm]{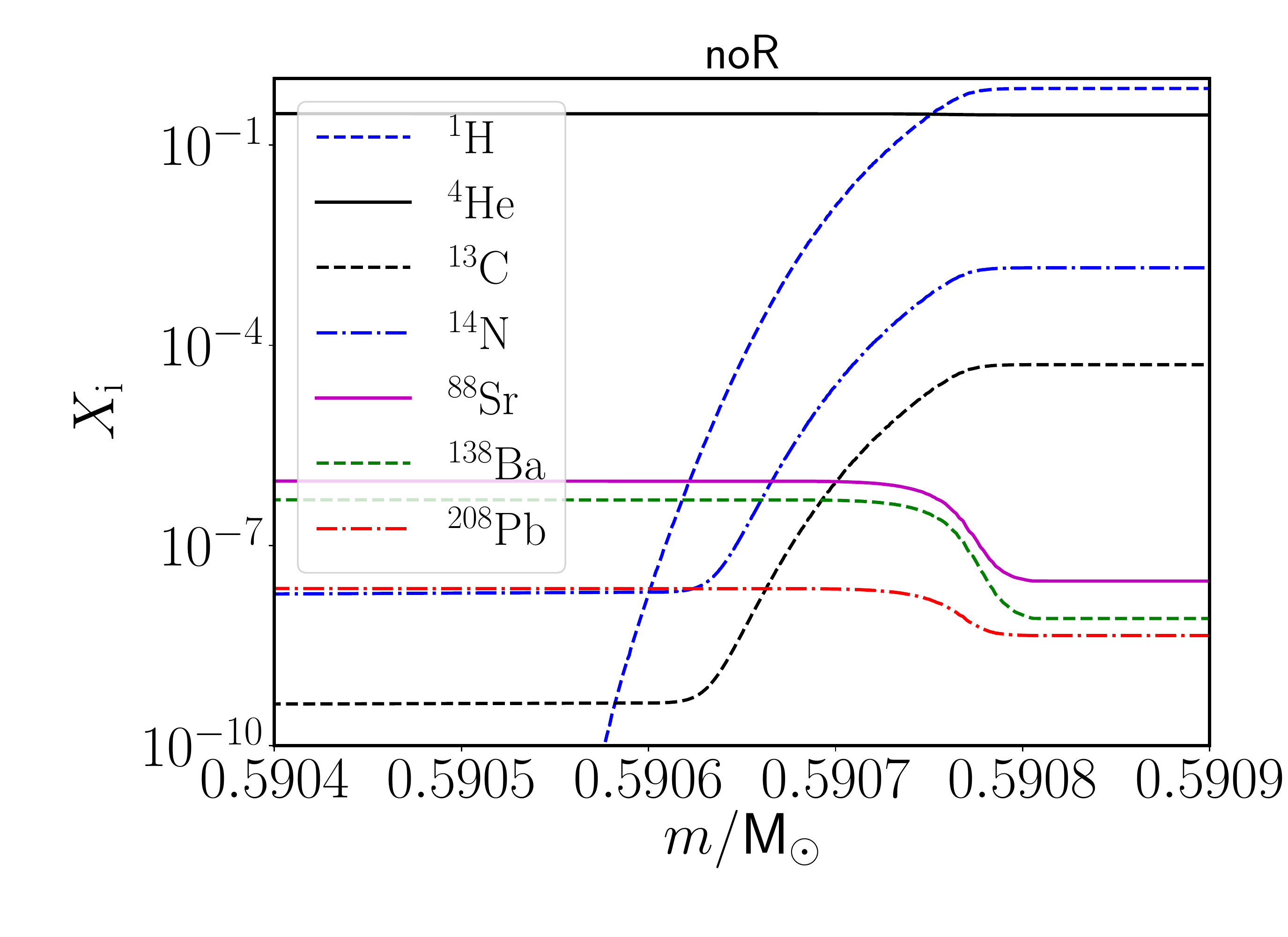}}}%
    \hspace{-0.13cm}
    \subfloat{{\includegraphics[width=6.15cm]{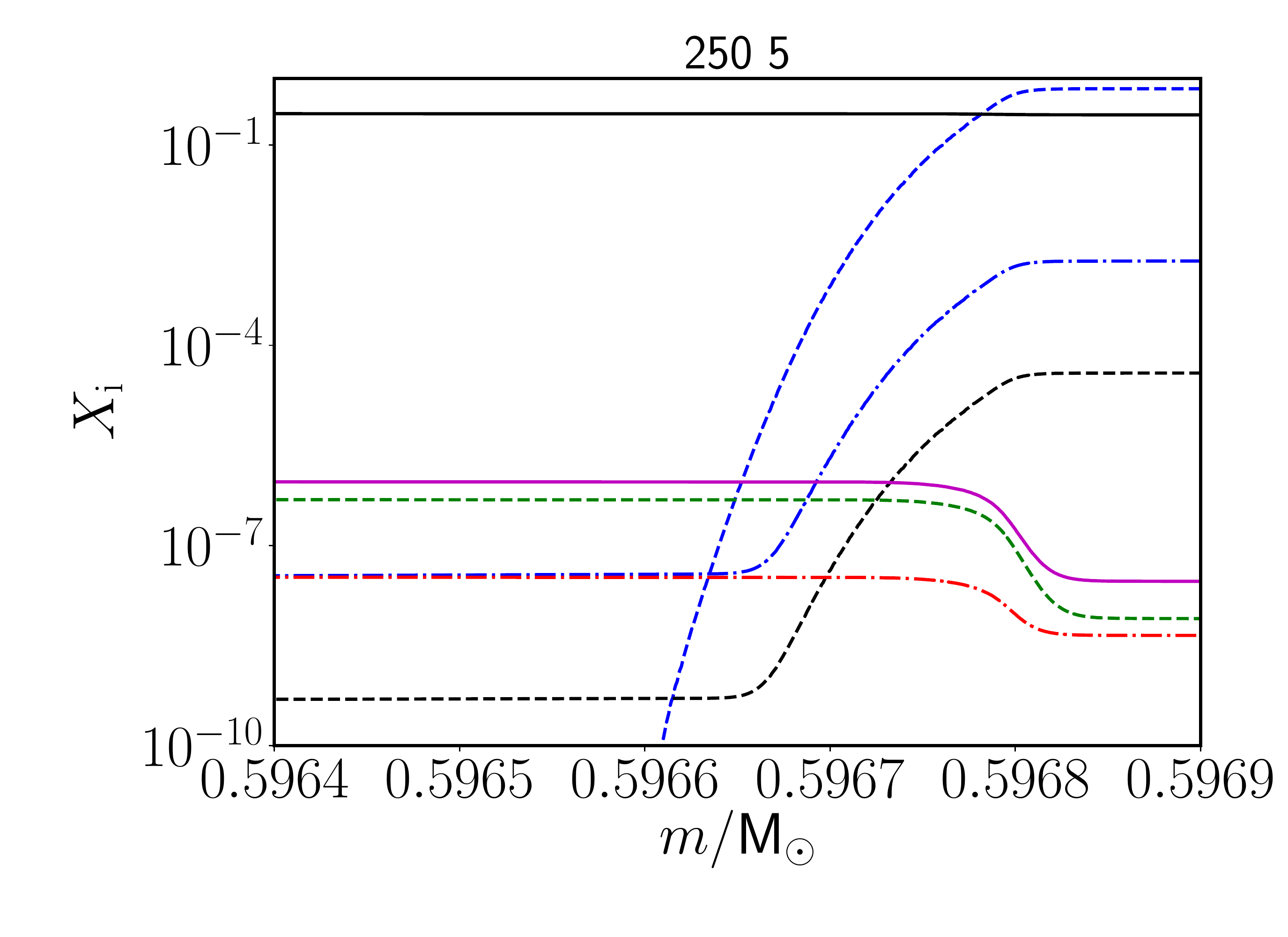}}}%
    \hspace{-0.17cm}
    \subfloat{{\includegraphics[width=6.15cm]{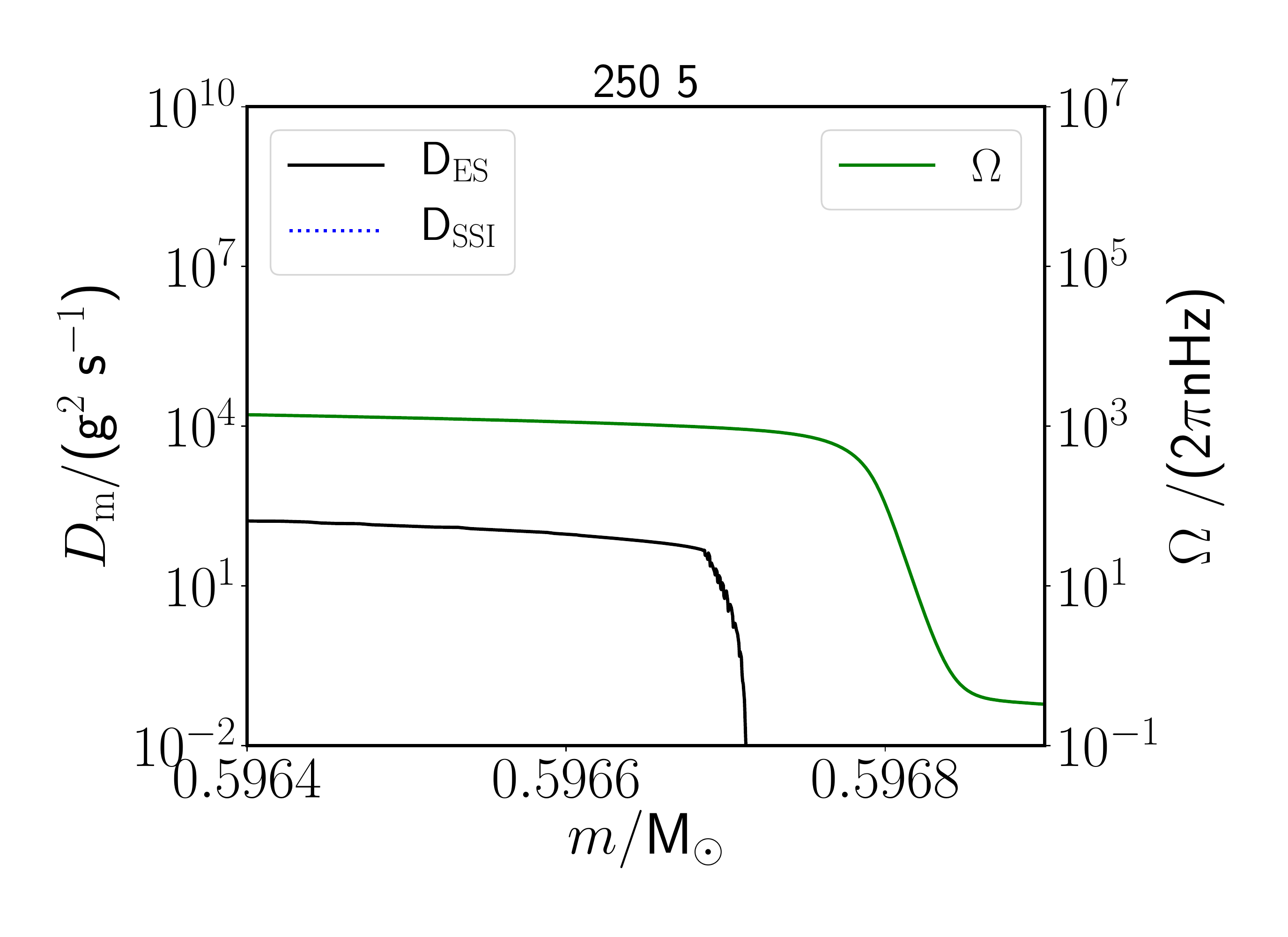}}}\\%
    \vspace{-1cm}
    \subfloat{{\includegraphics[width=6.15cm]{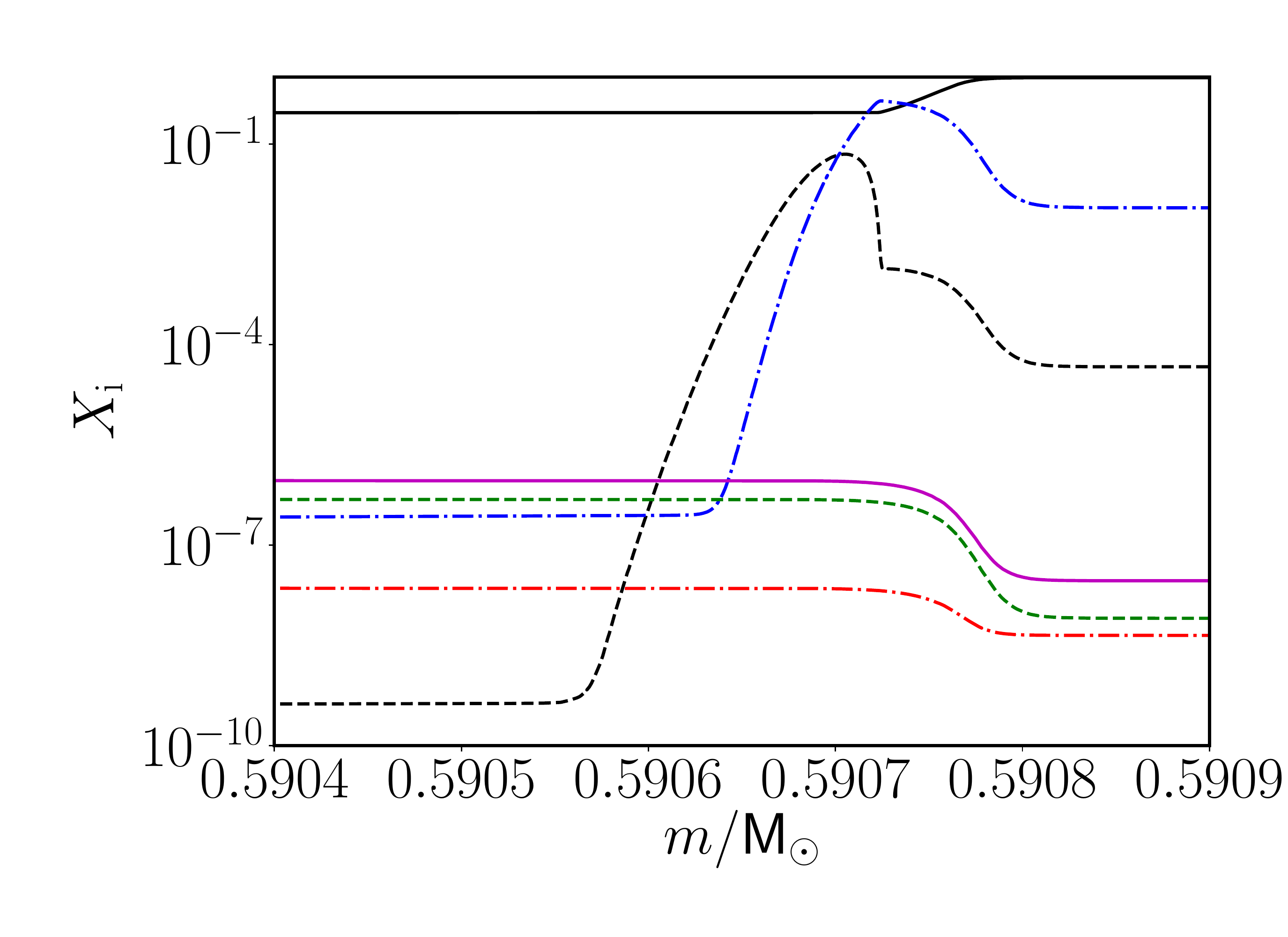}}}%
    \hspace{-0.13cm}
    \subfloat{{\includegraphics[width=6.15cm]{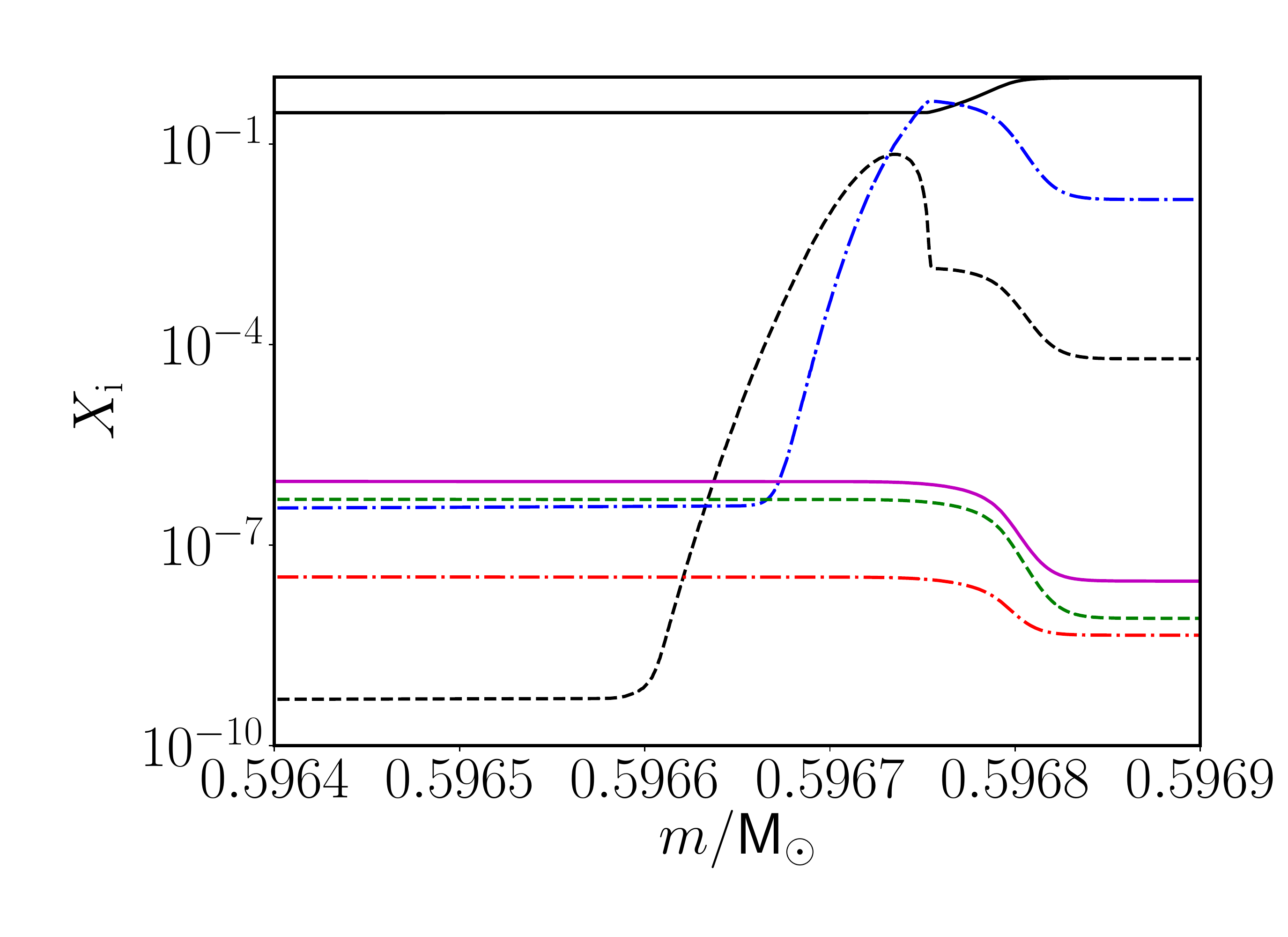}}}%
    \hspace{-0.17cm}
    \subfloat{{\includegraphics[width=6.15cm]{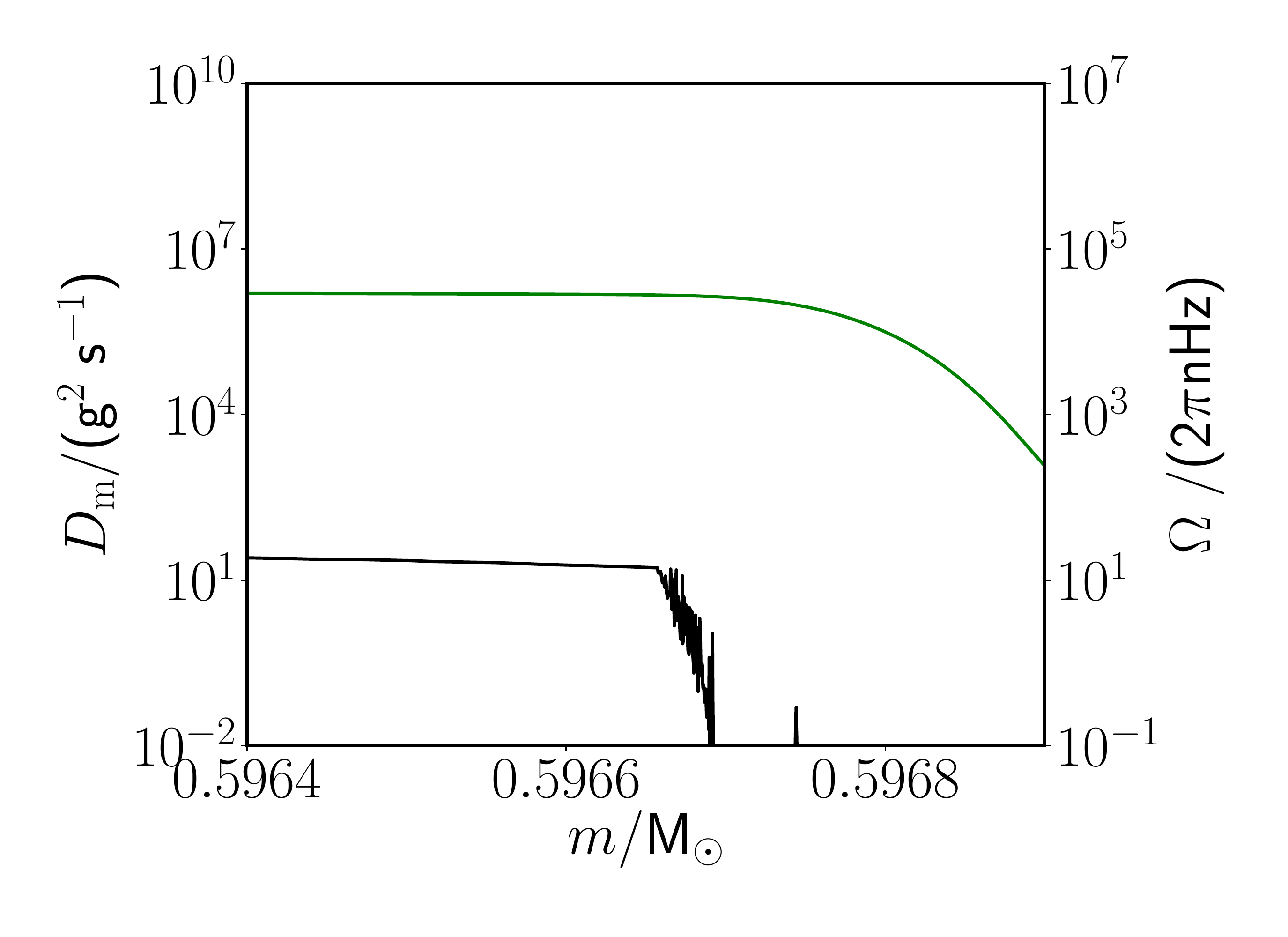}}}\\%
    \vspace{-1cm}
    \subfloat{{\includegraphics[width=6.15cm]{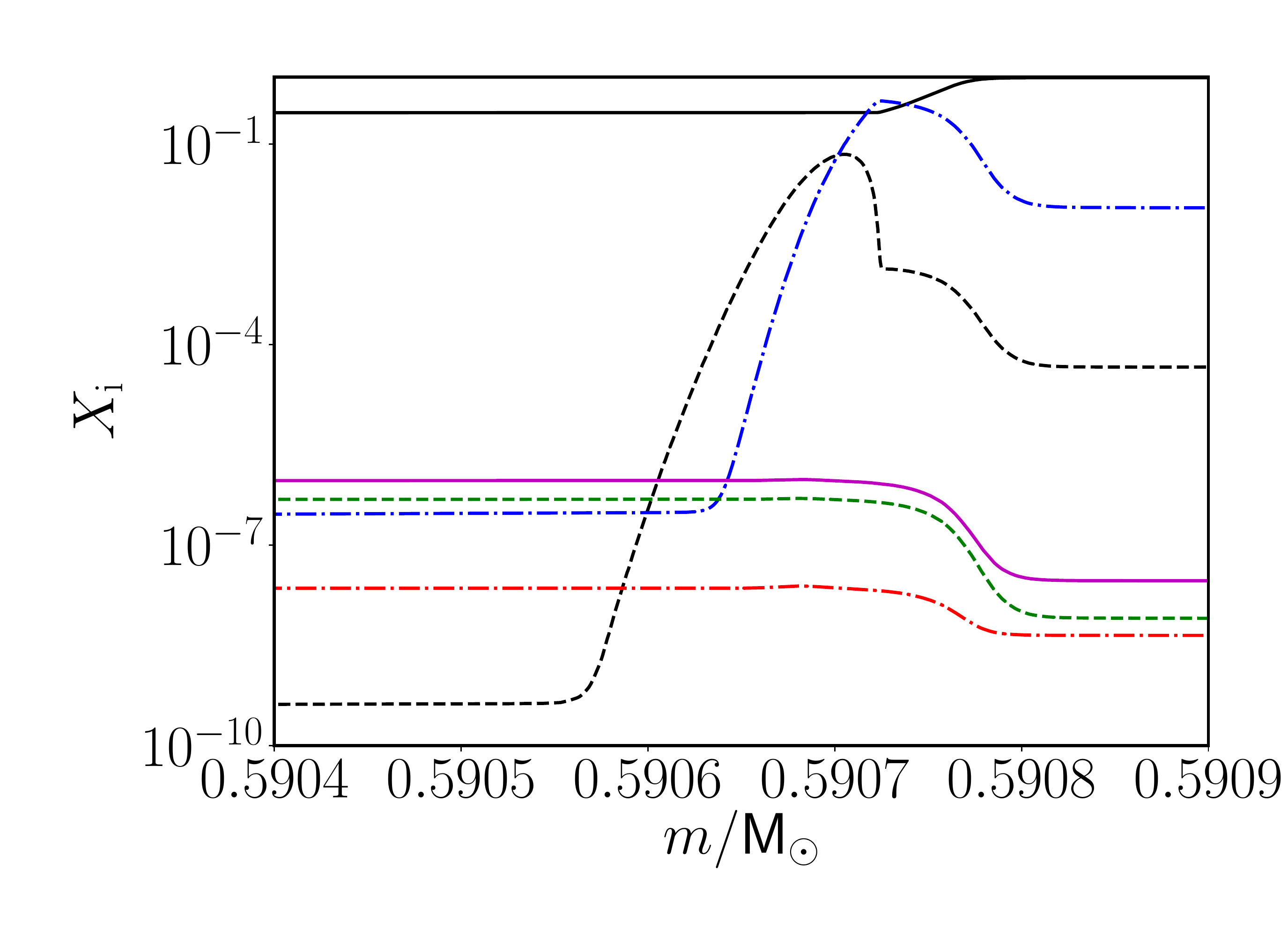}}}%
    \hspace{-0.13cm}
    \subfloat{{\includegraphics[width=6.15cm]{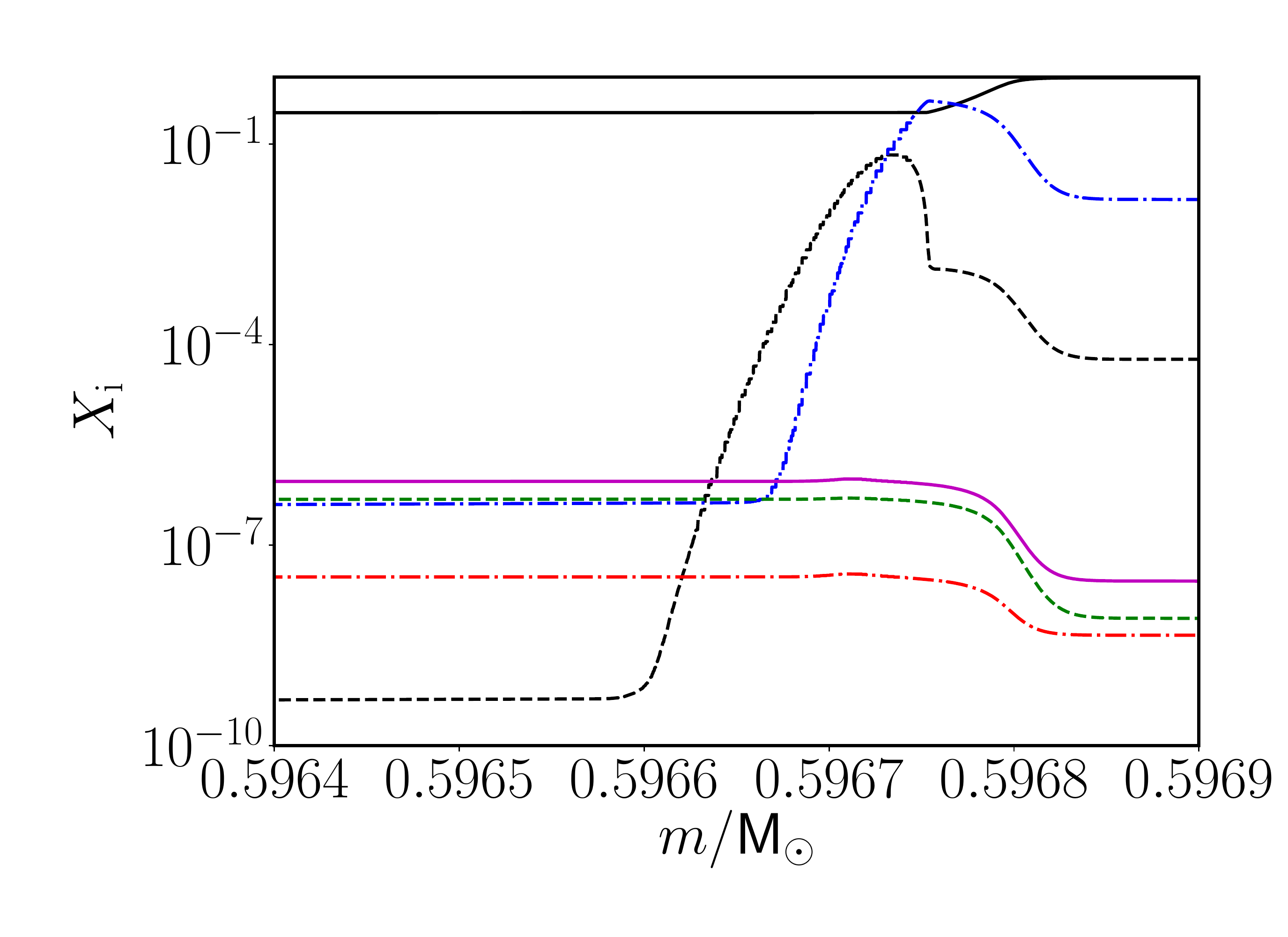}}}%
    \hspace{-0.17cm}
    \subfloat{{\includegraphics[width=6.15cm]{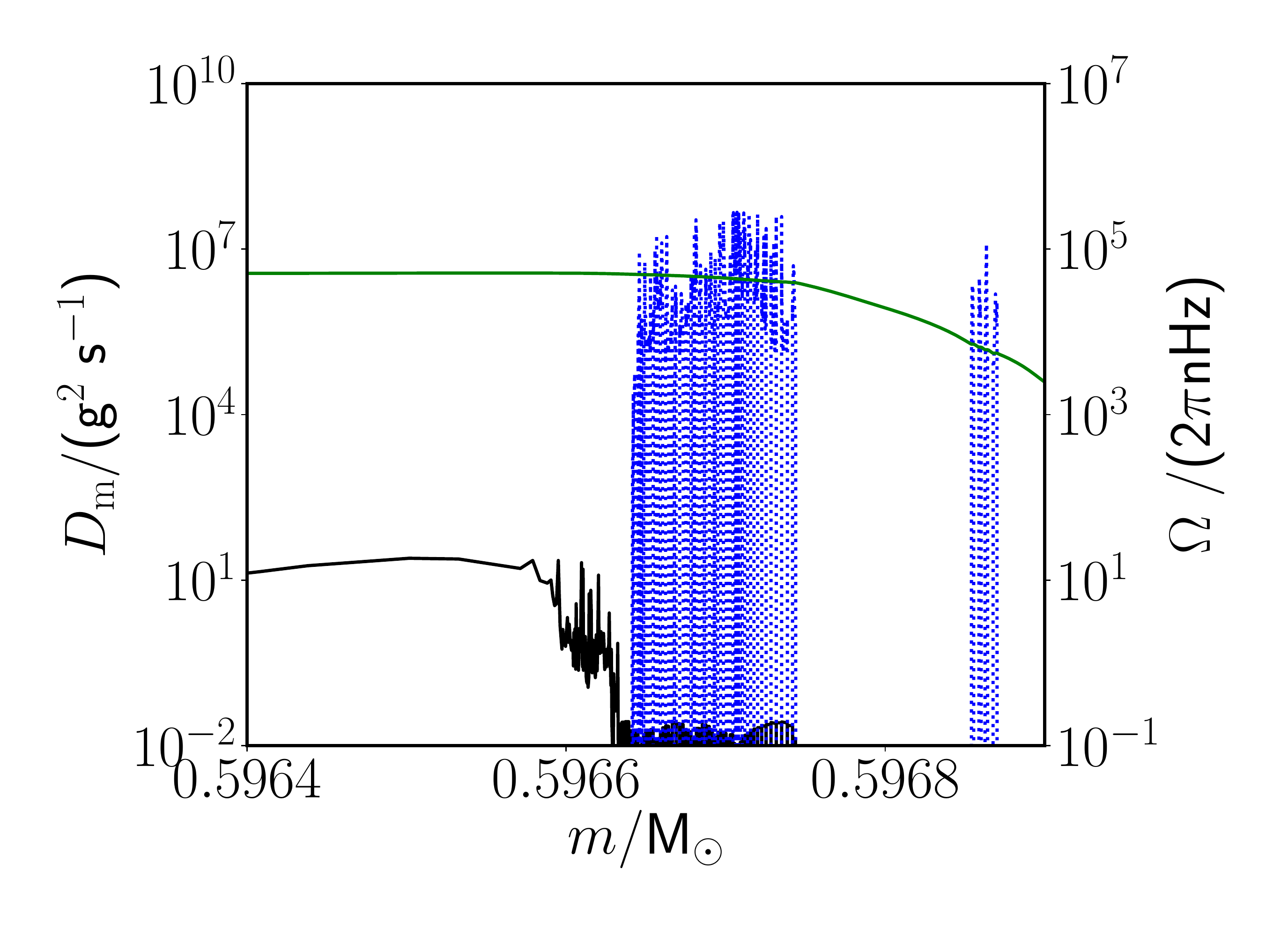}}}%
    \caption{Abundance and diffusion profiles within $^{13}$C-pocket regions. These regions fall within the same interpulse period as the fifth TDU (from our MPPNP results). The left panels show the abundance profiles of the non-rotating model, and the middle and right panels show the abundance and diffusion profiles of the `250 5' model. The top panels correspond to the maximum extent of the TDU, the middle panels correspond to the maximum $^{13}$C-pocket size, and the bottom panels correspond to the profiles when the s-process production has started. The influence of rotation on the $^{13}$C-pocket of the `250 5' is small, the only difference is that the abundance profiles are not as smooth as in the `noR' model.}%
    \label{fig:c13pckt_noR}%
\end{figure*}
In this section we show the s-process production of two the models from Table \ref{tab:mass}: the `250 5' and `250 6' models. We compare the s-process production of these models to the s-process production of our non-rotating model. The other models included in Table \ref{tab:mass}, which do not include an additional viscosity, are discussed in Appendix \ref{sec:app_sproc_noNUadd} together with a comparison to previously published work on s-process production in rotating AGB stars.

\subsection{$^{13}$C-pockets}
\label{sec:noR_pckt}
\begin{figure}
    \centering
    \includegraphics[width=8cm]{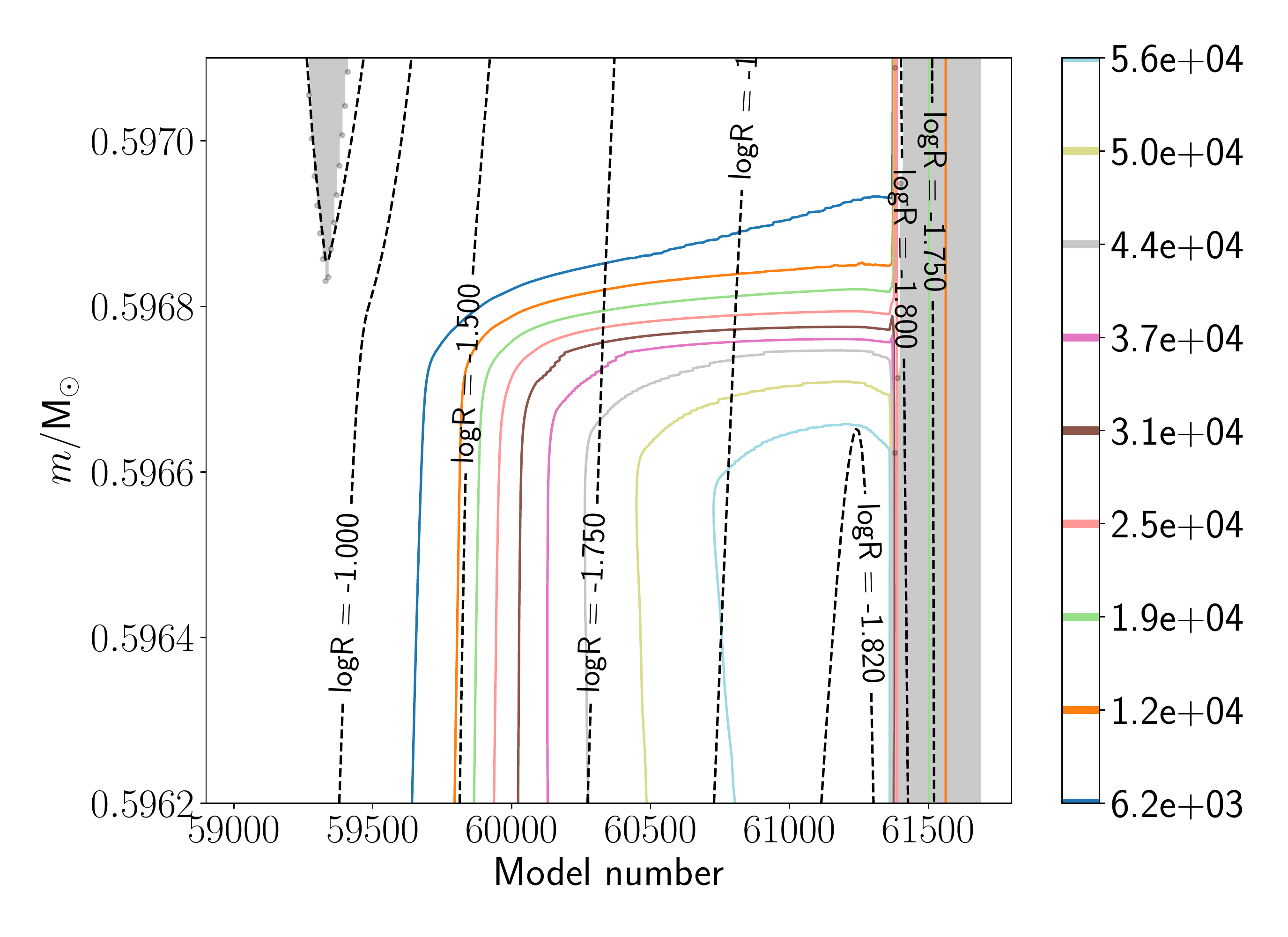}
    \caption{Time evolution of $\Omega$. The $\Omega$ profile is taken from the interpulse of the `250 5' model that is shown in Fig.\thinspace\ref{fig:c13pckt_noR}. Grey regions are the convective envelope during TDU (left) and the TP (right), dashed black contour lines show constant log$_{10}(r$/R$_{\odot}$), coloured contour lines show $\Omega$ values in linear range (the darker the contour line, the lower $\Omega$). Model numbers 59400, 60000, and 60500 correspond to the three time steps in Fig.\thinspace\ref{fig:c13pckt_noR}, and the vertical axis of this figure corresponds to the horizontal axes of the `250 5' panels in Fig.\thinspace\ref{fig:c13pckt_noR}. The contraction of the region leads to a steeper $\Omega$ gradient in the $^{13}$C-pocket region.}
    \label{fig:2D_omega}
\end{figure}

As explained in the Introduction, the $^{13}$C-pocket in low-mass AGB stars is where most of the neutrons for the neutron captures are produced. Therefore, we start our comparison with the abundance and diffusion profiles in the $^{13}$C-pockets. Specifically, we compare the $^{13}$C-pocket of the non-rotating and the `250 5' model during the interpulse period in which the fifth TDU takes place. We chose to use this model for this comparison as it will give us a conservative upper limit of the impact of rotation on the $^{13}$C-pocket.\\ 
The abundance and diffusion coefficient profiles of the $^{13}$C-pockets are shown in Fig.\thinspace\ref{fig:c13pckt_noR} for three different time steps. The diffusion profiles are calculated following \citet{2003ApJHerwig_rot}: we show the Lagrangian mixing coefficient ($D_{\mathrm{m}}$ and not the Eulerian one $D_{\mathrm{r}}$ which is given as MESA output) as we want to assess the effect of the mixing processes on the chemical elements:
\begin{equation}
D_{\mathrm{m}}=\left(\frac{dm}{dr}\right)^2 D_{\mathrm{r}}=(4\pi \rho r^2)^2D_{\mathrm{r}},
\label{eq:Ds}
\end{equation}
where all symbols have their usual meaning. In the same figure we also added the $\Omega$ profiles on log-scale, to better understand the behaviour of the instabilities. These $\Omega$-profiles show that the pocket is located just below the drop in $\Omega$, which coincides with the maximum extent of the TDU.\\
For all three time steps, the profiles and the size of the $^{13}$C-pocket in the two models are comparable, because the diffusion coefficient of the Eddington-Sweet (ES) circulation is present with values between 10$^{1}$-10$^{2}$ g$^2$ s$^{-1}$. This is not high enough to impact the abundance profiles. Also, the ES circulation is only present in regions of constant $\Omega$, which is also where the $^{13}$C abundance is low. The reason behind these characteristics can be explained by the strong dependence of $D_{\mathrm{ES}}$ on $\Omega$ \citep{2000ApJHeger}, which is $D_{\mathrm{m,ES}}\propto \Omega^2$. The $\Omega$ evolution during the interpulse phase of the `250 5' model is shown in Fig.\thinspace\ref{fig:2D_omega}. When $\Omega$ increases due to the contraction of the intershell region, $D_{\mathrm{m,ES}}$ remains nearly constant due to the smaller radial coordinate of the $^{13}$C-pocket. The ES-circulation is also dependent on the molecular weight gradient, which prevents this mixing process from being active within the $^{13}$C-pocket.

\begin{figure}[ht]
   \includegraphics[width=\linewidth]{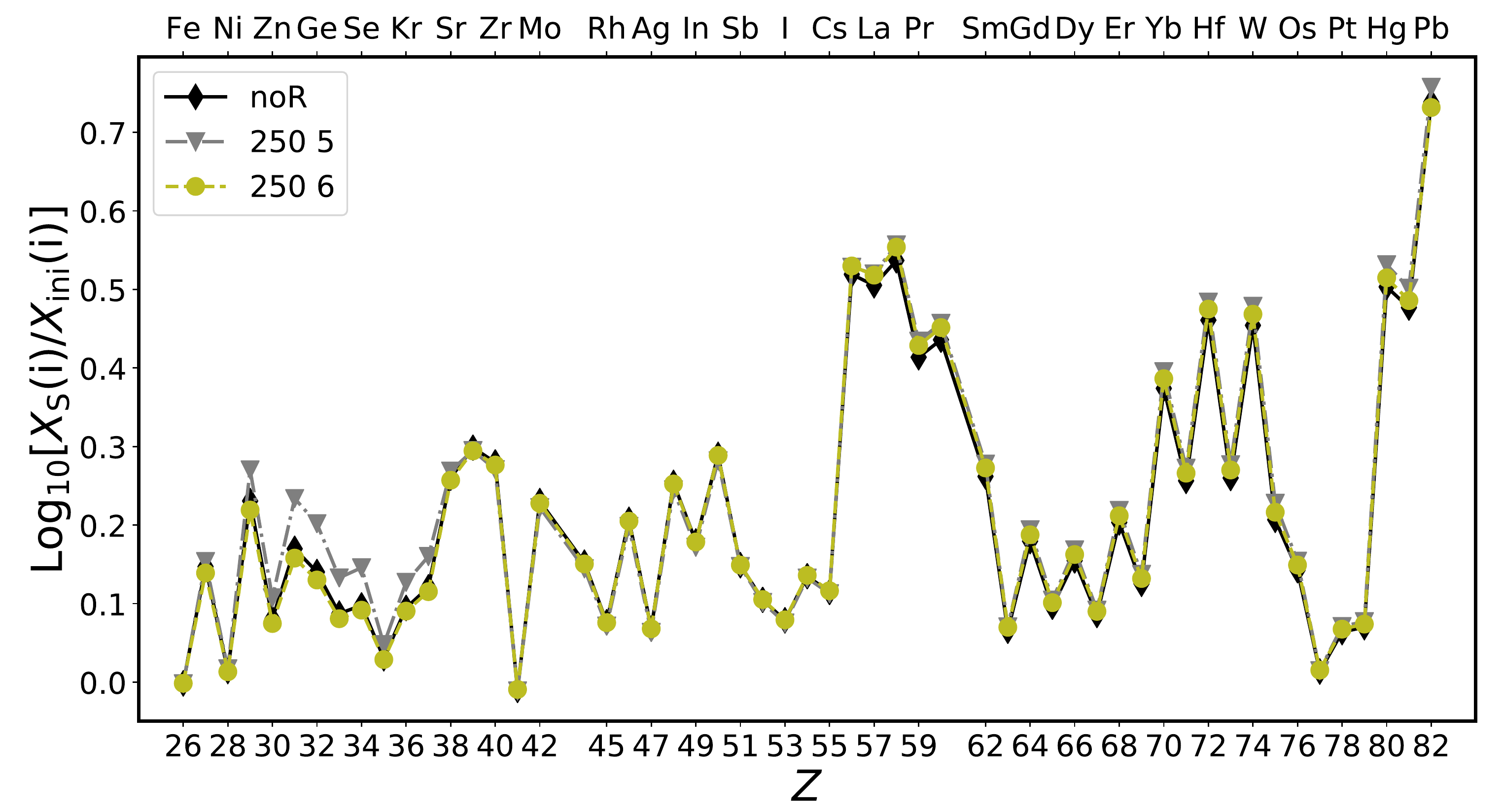}
   \caption{Surface enrichment of `noR', `250 5' and `250 6'. This comparison shows that the s-process production of rotating models that match asteroseismically measured rotation rates is comparable to that of the non-rotating model.}
   \label{fig:surf_enrich_slow}%
\end{figure}
\noindent The secular shear, the only other rotationally induced instability included in this model, is only present in the panels of Fig.\thinspace\ref{fig:c13pckt_noR} when the s-process production has started, in the region of the $^{13}$C-pocket. The $D_{\mathrm{m,SSI}}$ depends on d$\Omega$/d$r$, which is stronger in the bottom panel of Fig.\thinspace\ref{fig:c13pckt_noR}, as shown in Fig.\thinspace\ref{fig:2D_omega}. Molecular weight gradients inhibit $D_{\mathrm{m,SSI}}$, which is why $D_{\mathrm{m,SSI}}$ decreases around $m$/M$_{\odot}\simeq$0.59675. The high values of $D_{\mathrm{m,SSI}}$ however, have little effect on the abundance profiles as $D_{\mathrm{m,SSI}}$ is discontinuous (more details on this can be found in Appendix \ref{sec:app_sproc_noNUadd}). Continuous spatial mixing is needed to influence the abundance profiles and the resulting s-process production. It is unknown whether the discontinuous character of the SSI is physical or numerical \citep[see also][]{Aerts2018a}.\\
Diffusion coefficients of rotationally induced instabilities have been discussed in the previous publications on rotating AGB stars \citep[see][]{Langer1999,2003ApJHerwig_rot,2004Siess,2013_fruity_rotation}. These publications, however, discuss rotating models, that do not include a process able to decrease the core rotation rate in order to match the asteroseismically measured core rotation rates. Therefore, these models rotate too fast at the start of the AGB phase. This is clear from Column 5 in Table \ref{tab:mass}, where the standard rotating models `125 0' and `250 0' rotate three orders of magnitude faster than the models that match the asteroseismically measured core rotation rates (`125 6' and `250 6'). Therefore, a consistent comparison is not possible between the models of previous publications and models described in this section. Here we only note that Fig.\thinspace2 of \citet{2013_fruity_rotation} shows the location in the intershell where their models are unstable against the ES circulation and the GSF instability (not present in our models, as discussed in Sect.\thinspace\ref{sec:rot_sett}. While \citet{2013_fruity_rotation} does not mention the strength of their diffusion coefficients, their Fig.\thinspace2 shows that they found the interpulse to be unstable for ES circulation at the same location as in our models.

\subsection{Surface enrichment of s-process elements}
In the previous subsection we found that rotation results only in small differences in the $^{13}$C-pockets when the `250 5' model is compared to the non rotating model. We therefore expect the resulting s-process production of the two models to be comparable. \\
In Fig.\thinspace\ref{fig:surf_enrich_slow} we show the surface enrichment factors for the models `noR', `250 5', and `250 6'. The surface enrichment factors have been calculated after the final TDU and are scaled to their initial abundances. All three models largely overlap in this figure.  The `noR' model experienced one TDU more than the two rotation models, we therefore show the surface enrichment of the TDU before the last TDU for the `noR' model to have a fair comparison.\\
From this we conclude that when the models rotate at a rate that matches the asteroseismically measured rotation rates or an order of magnitude faster, the s-process production is comparable to that of the non-rotating model, as suggested by \citet{2013_fruity_rotation}. A consequence of this result is that, according to our models, any spread in observed s-process production of a certain metallicity is unlikely to be caused by rotation (see e.g. \citealt{2002Abia} and \citealt{DeCastro2016}).\\
As the results of the rotating models described in this section match the non-rotating model, we refer the reader to \citet{Battino_2016} and \citet{2019Battino} for a comparison to s-process observations, because the non-rotating models described in that paper are similar to those presented here. \\


\section{Final remarks}
\label{sec:concl}
In this paper we presented rotating AGB star models (2 M$_{\odot}$, Z=0.01) that are enforced to match the asteroseismically measured rotation rates before and after the AGB phase. For the first time, we have presented the s-process production of such models that rotate at such rates. Our main findings are described below.\\
\begin{itemize}
\item Our models including additional viscosity of $\nu_{\rm{add}}$= 10$^6$ cm$^2$ s$^{-1}$ follow the upper limit of the observed trend of core and envelope rotation rates inferred from Kepler observations, comparable to the results of \citet{Jacqueline1}. 
\item The models that are enforced to match the asteroseismically measured core rotation rate show s-process production similar to that of the non-rotating model. Therefore the effect of rotation on s-process production is negligible in these models. 
\item We also calculated a model where the core rotates an order of magnitude faster than observed values, as conservative upper limit to observed rotation rates. The s-process production of this model is also comparable to the non-rotating model, strengthening our previous conclusion.
\item The results above are independent of the initial rotation rate.
\end{itemize}
Several uncertainties may potentially affect these conclusions. The most important is the constant $\nu_{\mathrm{add}}$ that is used to reduce the theoretical core rotation rates to the asteroseismically obtained rates. This constant has no physical meaning (yet) and the results presented here should therefore be interpreted as not necessarily the final answer, but as a next step towards understanding the s-process production in rotating low-mass AGB stars. In particular, different combinations of the value for $\nu_{\mathrm{add}}$ and the values of the two $f$ parameters in the implementation of rotation may lead to similar core rotation rates. The range of values for these $f$ parameters might however be limited, as more recent calibrations by \citet{2006Yoon} and \citet{2011Brott} resulted in $f_{\mu}$=0.1, $f_{\mathrm{c}}$=0.03), similar values to the ones found by \citet{2000ApJHeger}. Our conclusions remain the same when we tested these values in our calculations.\\
Another caveat to counter is that the missing process of angular momentum could also mix chemical elements. When we include a $\nu_{\mathrm{add}}$ into the mixing of chemical elements with same value as for the $\nu_{\mathrm{add}}$, we find chemically homogeneously evolving stars \citep[as first described by][]{1987Maeder}. In the low-mass regime, there is no observational evidence for these stars and we therefore infer that the missing process of angular momentum cannot have the same efficiency for both the transport of angular momentum as for the mixing of chemical elements.
Another point is that we have only investigated the effects of non-magnetic mixing processes. In Paper I, we already found that the TS-dynamo does not allow for enough transport of angular momentum in stellar evolutionary models with an initial mass of 2.5 M$_{\odot}$ to match the observations \citep[confirming results of][for their 1.5 M$_{\odot}$ models]{2014cantiello}. Recently, a revised derivation of the TS-dynamo was published by \citet{2019Fuller} who show that this mechanism is able to match the asteroseismically obtained core rotation rates. However, this prescription is unable to match the rotational profile of the Sun \citep{2019Eggenberger}.\\
Besides the uncertainties around the missing process of angular momentum transport, the current implementation of rotationally induced mixing processes remains a major challenge (Appendix \ref{sec:app_sproc_noNUadd}). We cannot exclude the possibility that better descriptions will effect the s-process production in rotating AGB stars. Furthermore, two flavours for the implementation of rotation in stellar evolution codes exist: diffusive \citep[see e.g.][]{2000ApJHeger} and advective \citep[see e.g.][]{2000ARA&AMandM,ReviewMandM}, where the second implementation uses different prescriptions for the mixing processes \textbf{and this could affect the s-process production in AGB stars.} \\
We will investigate these uncertainties in future publications.

\begin{acknowledgements}
This work has been supported by the European Research Council (ERC-2012-St Grant 306901, ERC-2015-STG Nr. 677497, ERC-2016-CO Grant 724560), and the EU COST Action CA16117 (ChETEC). This work is part of the BRIDGCE UK Network is a UK-wide network established to Bridge Research in the different Disciplines related to the Galactic Chemical Evolution and nuclear astrophysics. We acknowledge significant support to NuGrid from NSF grant PHY-1430152 (JINA Center for the Evolution of the Elements) and STFC (through the University of Hull's Consolidated Grant ST/R000840/1). RH acknowledges support from the World Premier International Research Centre Initiative (WPI Initiative), MEXT, Japan. We performed the MPPNP calculations on the Viper High Performance Computing facility of the University of Hull and acknowledge its support team. We thank the referee for her/his detailed comments that helped improve this paper.
\end{acknowledgements}

\bibliographystyle{aa} 
\bibliography{references.bib} 

\begin{thebibliography}{85}
\expandafter\ifx\csname natexlab\endcsname\relax\def\natexlab#1{#1}\fi

\bibitem[{{Abia} {et~al.}(2002){Abia}, {Dom{\'{\i}}nguez}, {Gallino}, {Busso},
  {Masera}, {Straniero}, {de Laverny}, {Plez}, \& {Isern}}]{2002Abia}
{Abia}, C., {Dom{\'{\i}}nguez}, I., {Gallino}, R., {et~al.} 2002, \apj, 579,
  817

\bibitem[{Aerts {et~al.}(2019)Aerts, Mathis, \& Rogers}]{Aerts2018}
Aerts, C., Mathis, S., \& Rogers, T.~M. 2019, \araa, 57, null

\bibitem[{{Aerts} {et~al.}(2018){Aerts}, {Molenberghs}, {Michielsen},
  {Pedersen}, {Bj{\"o}rklund}, {Johnston}, {Mombarg}, {Bowman}, {Buysschaert},
  {P{\'a}pics}, {Sekaran}, {Sundqvist}, {Tkachenko}, {Truyaert}, {Van Reeth},
  \& {Vermeyen}}]{Aerts2018a}
{Aerts}, C., {Molenberghs}, G., {Michielsen}, M., {et~al.} 2018, \apjs, 237, 15

\bibitem[{Aerts {et~al.}(2017)Aerts, Reeth, \& Tkachenko}]{Aerts2017}
Aerts, C., Reeth, T.~V., \& Tkachenko, A. 2017, \apj, 847, L7

\bibitem[{Battino {et~al.}(2016)Battino, Pignatari, Ritter, Herwig, Denisenkov,
  Hartogh, Trappitsch, Hirschi, Freytag, Thielemann, \& Paxton}]{Battino_2016}
Battino, U., Pignatari, M., Ritter, C., {et~al.} 2016, \apj, 827, 30

\bibitem[{{Battino} {et~al.}(2019){Battino}, {Tattersall}, {Lederer-Woods},
  {Herwig}, {Denissenkov}, {Hirschi}, {Trappitsch}, {den Hartogh}, \&
  {Pignatari}}]{2019Battino}
{Battino}, U., {Tattersall}, A., {Lederer-Woods}, C., {et~al.} 2019, arXiv
  e-prints, arXiv:1906.01952

\bibitem[{{Bennett} {et~al.}(2012){Bennett}, {Hirschi}, {Pignatari}, {Diehl},
  {Fryer}, {Herwig}, {Hungerford}, {Nomoto}, {Rockefeller}, {Timmes}, \&
  {Wiescher}}]{Bennett2012}
{Bennett}, M.~E., {Hirschi}, R., {Pignatari}, M., {et~al.} 2012, \mnras, 420,
  3047

\bibitem[{Bisterzo {et~al.}(2017)Bisterzo, Travaglio, Wiescher, Käppeler, \&
  Gallino}]{Bisterzo2017}
Bisterzo, S., Travaglio, C., Wiescher, M., Käppeler, F., \& Gallino, R. 2017,
  \apj, 835, 97

\bibitem[{{Bloecker}(1995)}]{1995blocker}
{Bloecker}, T. 1995, \aap, 297, 727

\bibitem[{{Borucki} {et~al.}(2010){Borucki}, {Koch}, {Basri}, {Batalha},
  {Brown}, {Caldwell}, {Caldwell}, {Christensen-Dalsgaard}, {Cochran},
  {DeVore}, {Dunham}, {Dupree}, {Gautier}, {Geary}, {Gilliland}, {Gould},
  {Howell}, {Jenkins}, {Kondo}, {Latham}, {Marcy}, {Meibom}, {Kjeldsen},
  {Lissauer}, {Monet}, {Morrison}, {Sasselov}, {Tarter}, {Boss}, {Brownlee},
  {Owen}, {Buzasi}, {Charbonneau}, {Doyle}, {Fortney}, {Ford}, {Holman},
  {Seager}, {Steffen}, {Welsh}, {Rowe}, {Anderson}, {Buchhave}, {Ciardi},
  {Walkowicz}, {Sherry}, {Horch}, {Isaacson}, {Everett}, {Fischer}, {Torres},
  {Johnson}, {Endl}, {MacQueen}, {Bryson}, {Dotson}, {Haas}, {Kolodziejczak},
  {Van Cleve}, {Chandrasekaran}, {Twicken}, {Quintana}, {Clarke}, {Allen},
  {Li}, {Wu}, {Tenenbaum}, {Verner}, {Bruhweiler}, {Barnes}, \&
  {Prsa}}]{2010borucki}
{Borucki}, W.~J., {Koch}, D., {Basri}, G., {et~al.} 2010, Science, 327, 977

\bibitem[{{Brott} {et~al.}(2011){Brott}, {de Mink}, {Cantiello}, {Langer}, {de
  Koter}, {Evans}, {Hunter}, {Trundle}, \& {Vink}}]{2011Brott}
{Brott}, I., {de Mink}, S.~E., {Cantiello}, M., {et~al.} 2011, \aap, 530, A115

\bibitem[{{Buntain} {et~al.}(2017){Buntain}, {Doherty}, {Lugaro}, {Lattanzio},
  {Stancliffe}, \& {Karakas}}]{Buntain2017}
{Buntain}, J.~F., {Doherty}, C.~L., {Lugaro}, M., {et~al.} 2017, \mnras, 471,
  824

\bibitem[{Burbidge {et~al.}(1957)Burbidge, Burbidge, Fowler, \& Hoyle}]{B2FH}
Burbidge, E.~M., Burbidge, G.~R., Fowler, W.~A., \& Hoyle, F. 1957, Rev. Mod.
  Phys., 29, 547

\bibitem[{{Caleo} {et~al.}(2016){Caleo}, {Balbus}, \& {Tognelli}}]{2016caleo}
{Caleo}, A., {Balbus}, S.~A., \& {Tognelli}, E. 2016, \mnras, 460, 338

\bibitem[{{Cantiello} {et~al.}(2014){Cantiello}, {Mankovich}, {Bildsten},
  {Christensen-Dalsgaard}, \& {Paxton}}]{2014cantiello}
{Cantiello}, M., {Mankovich}, C., {Bildsten}, L., {Christensen-Dalsgaard}, J.,
  \& {Paxton}, B. 2014, \apj, 788, 93

\bibitem[{{Ceillier} {et~al.}(2017){Ceillier}, {Tayar}, {Mathur}, {Salabert},
  {Garc{\'{\i}}a}, {Stello}, {Pinsonneault}, {van Saders}, {Beck}, \&
  {Bloemen}}]{Ceillier2017}
{Ceillier}, T., {Tayar}, J., {Mathur}, S., {et~al.} 2017, \aap, 605, A111

\bibitem[{{Chaboyer} {et~al.}(1995){Chaboyer}, {Demarque}, \&
  {Pinsonneault}}]{cha95}
{Chaboyer}, B., {Demarque}, P., \& {Pinsonneault}, M.~H. 1995, \apj, 441, 865

\bibitem[{{Chaboyer} \& {Zahn}(1992)}]{1992chaboyer}
{Chaboyer}, B. \& {Zahn}, J.-P. 1992, \aap, 253, 173

\bibitem[{{de Castro} {et~al.}(2016){de Castro}, {Pereira}, {Roig}, {Jilinski},
  {Drake}, {Chavero}, \& {Sales Silva}}]{DeCastro2016}
{de Castro}, D.~B., {Pereira}, C.~B., {Roig}, F., {et~al.} 2016, \mnras, 459,
  4299

\bibitem[{Deheuvels {et~al.}(2015)Deheuvels, Ballot, Beck, \&
  Mosser}]{Deheuvels2015}
Deheuvels, S., Ballot, J., Beck, P., \& Mosser, B. 2015, A\&A, 580

\bibitem[{Deheuvels {et~al.}(2014)Deheuvels, Do{\u{g}}an, Goupil, Appourchaux,
  Benomar, Bruntt, Campante, Casagrande, Ceillier, Davies, Cat, Fu,
  Garc{\'{\i}}a, Lobel, Mosser, Reese, Regulo, Schou, Stahn, Thygesen, Yang,
  Chaplin, Christensen-Dalsgaard, Eggenberger, Gizon, Mathis,
  Molenda-{\.{Z}}akowicz, \& Pinsonneault}]{Deheuvels2014}
Deheuvels, S., Do{\u{g}}an, G., Goupil, M.~J., {et~al.} 2014, \aap, 564, A27

\bibitem[{Deheuvels {et~al.}(2012)Deheuvels, Garc{\'{\i}}a, Chaplin, Basu,
  Antia, Appourchaux, Benomar, Davies, Elsworth, Gizon, Goupil, Reese, Regulo,
  Schou, Stahn, Casagrande, Christensen-Dalsgaard, Fischer, Hekker, Kjeldsen,
  Mathur, Mosser, Pinsonneault, Valenti, Christiansen, Kinemuchi, \&
  Mullally}]{2012deheuvels}
Deheuvels, S., Garc{\'{\i}}a, R.~A., Chaplin, W.~J., {et~al.} 2012, \apj, 756,
  19

\bibitem[{{den Hartogh} {et~al.}(2019){den Hartogh}, {Eggenberger}, \&
  {Hirschi}}]{Jacqueline1}
{den Hartogh}, J.~W., {Eggenberger}, P., \& {Hirschi}, R. 2019, \aap, 622, A187

\bibitem[{Denissenkov {et~al.}(2010)Denissenkov, Pinsonneault, Terndrup, \&
  Newsham}]{Denissenkov2010}
Denissenkov, P.~A., Pinsonneault, M., Terndrup, D.~M., \& Newsham, G. 2010,
  ApJ, 716, 1269

\bibitem[{{Denissenkov} \& {Tout}(2003)}]{2003Denissenkov}
{Denissenkov}, P.~A. \& {Tout}, C.~A. 2003, \mnras, 340, 722

\bibitem[{Eddington(1925)}]{Eddington1925}
Eddington, A.~S. 1925, The Observatory, 48, 73

\bibitem[{{Eggenberger} {et~al.}(2019){Eggenberger}, {Buldgen}, \&
  {Salmon}}]{2019Eggenberger}
{Eggenberger}, P., {Buldgen}, G., \& {Salmon}, S.~J.~A.~J. 2019, \aap, 626, L1

\bibitem[{Eggenberger {et~al.}(2019)Eggenberger, Deheuvels, Miglio,
  Ekstr{\"o}m, Georgy, Meynet, N., Salmon, Buldgen, Montalb{\'a}n, Spada, \&
  Ballot}]{Eggenberger2019}
Eggenberger, P., Deheuvels, S., Miglio, A., {et~al.} 2019, \aap, 621, A66

\bibitem[{Eggenberger {et~al.}(2017)Eggenberger, Lagarde, Miglio,
  Montalb{\'{a}}n, Ekström, Georgy, Meynet, Salmon, Ceillier, Garc{\'{\i}}a,
  Mathis, Deheuvels, Maeder, den Hartogh, \& Hirschi}]{Eggenberger_2017}
Eggenberger, P., Lagarde, N., Miglio, A., {et~al.} 2017, \aap, 599, A18

\bibitem[{{Eggenberger} {et~al.}(2005){Eggenberger}, {Maeder}, \&
  {Meynet}}]{egg05}
{Eggenberger}, P., {Maeder}, A., \& {Meynet}, G. 2005, \aap, 440, L9

\bibitem[{Eggenberger {et~al.}(2012)Eggenberger, Montalb{\'a}n, \&
  Miglio}]{2012eggenberger}
Eggenberger, P., Montalb{\'a}n, J., \& Miglio, A. 2012, \aap, 544, L4

\bibitem[{{Endal} \& {Sofia}(1978)}]{endal_sofia1978}
{Endal}, A.~S. \& {Sofia}, S. 1978, \apj, 220, 279

\bibitem[{{Fricke}(1968)}]{Fricke1968}
{Fricke}, K. 1968, \zap, 68, 317

\bibitem[{{Fuller} {et~al.}(2019){Fuller}, {Piro}, \& {Jermyn}}]{2019Fuller}
{Fuller}, J., {Piro}, A.~L., \& {Jermyn}, A.~S. 2019, \mnras, 485, 3661

\bibitem[{{Gallino} {et~al.}(1998){Gallino}, {Arlandini}, {Busso}, {Lugaro},
  {Travaglio}, {Straniero}, {Chieffi}, \& {Limongi}}]{1998Gallino}
{Gallino}, R., {Arlandini}, C., {Busso}, M., {et~al.} 1998, \apj, 497, 388

\bibitem[{{Goldreich} \& {Schubert}(1967)}]{Goldreich1967}
{Goldreich}, P. \& {Schubert}, G. 1967, \apj, 150, 571

\bibitem[{Goriely \& Siess(2018)}]{Goriely2017}
Goriely, S. \& Siess, L. 2018, \aap, 609, A29

\bibitem[{{Grevesse} \& {Noels}(1993)}]{1993grevesse}
{Grevesse}, N. \& {Noels}, A. 1993, in Perfectionnement de l'Association
  Vaudoise des Chercheurs en Physique, ed. B.~{Hauck}, S.~{Paltani}, \&
  D.~{Raboud}, 205--257

\bibitem[{{Heger} {et~al.}(2000){Heger}, {Langer}, \& {Woosley}}]{2000ApJHeger}
{Heger}, A., {Langer}, N., \& {Woosley}, S.~E. 2000, \apj, 528, 368

\bibitem[{Hermes {et~al.}(2017)Hermes, Gänsicke, Kawaler, Greiss, Tremblay,
  Fusillo, Raddi, Fanale, Bell, Dennihy, Fuchs, Dunlap, Clemens, Montgomery,
  Winget, Chote, Marsh, \& Redfield}]{Hermes2017}
Hermes, J.~J., Gänsicke, B.~T., Kawaler, S.~D., {et~al.} 2017, The
  Astrophysical Journal Supplement Series, 232, 23

\bibitem[{{Herwig}(2000)}]{2000herwig}
{Herwig}, F. 2000, \aap, 360, 952

\bibitem[{Herwig(2001)}]{Herwig1999}
Herwig, F. 2001, Astrophysics and Space Science, 275, 15

\bibitem[{{Herwig}(2005)}]{falk_ARAA}
{Herwig}, F. 2005, \araa, 43, 435

\bibitem[{{Herwig} {et~al.}(1997){Herwig}, {Bloecker}, {Schoenberner}, \& {El
  Eid}}]{1997herwig_overs}
{Herwig}, F., {Bloecker}, T., {Schoenberner}, D., \& {El Eid}, M. 1997, \aap,
  324, L81

\bibitem[{{Herwig} {et~al.}(2007){Herwig}, {Freytag}, {Fuchs}, {Hansen},
  {Hueckstaedt}, {Porter}, {Timmes}, \& {Woodward}}]{2007herwig}
{Herwig}, F., {Freytag}, B., {Fuchs}, T., {et~al.} 2007, in Astronomical
  Society of the Pacific Conference Series, Vol. 378, Why Galaxies Care About
  AGB Stars: Their Importance as Actors and Probes, ed. F.~{Kerschbaum},
  C.~{Charbonnel}, \& R.~F. {Wing}, 43

\bibitem[{{Herwig} {et~al.}(2003){Herwig}, {Langer}, \&
  {Lugaro}}]{2003ApJHerwig_rot}
{Herwig}, F., {Langer}, N., \& {Lugaro}, M. 2003, \apj, 593, 1056

\bibitem[{{Hirschi} \& {Maeder}(2010)}]{2010Hirschi}
{Hirschi}, R. \& {Maeder}, A. 2010, \aap, 519, A16

\bibitem[{{Hollowell} \& {Iben}(1988)}]{1988Hollowell}
{Hollowell}, D. \& {Iben}, Icko, J. 1988, \apjl, 333, L25

\bibitem[{{Huang} {et~al.}(2010){Huang}, {Gies}, \& {McSwain}}]{2010Huang}
{Huang}, W., {Gies}, D.~R., \& {McSwain}, M.~V. 2010, \apj, 722, 605

\bibitem[{{Iben} \& {Renzini}(1982)}]{1982IbenRenzini}
{Iben}, I., J. \& {Renzini}, A. 1982, \apjl, 263, L23

\bibitem[{{James} \& {Kahn}(1970)}]{1970James}
{James}, H.~A. \& {Kahn}, F.~D. 1970, \aap, 5, 232

\bibitem[{{James} \& {Kahn}(1971)}]{1971James}
{James}, R.~A. \& {Kahn}, F.~D. 1971, \aap, 12, 332

\bibitem[{K\"appeler {et~al.}(2011)K\"appeler, Gallino, Bisterzo, \&
  Aoki}]{Kappeler2011}
K\"appeler, F., Gallino, R., Bisterzo, S., \& Aoki, W. 2011, Rev. Mod. Phys.,
  83, 157

\bibitem[{{Karakas} \& {Lattanzio}(2014)}]{2014PASA}
{Karakas}, A.~I. \& {Lattanzio}, J.~C. 2014, \pasa, 31, e030

\bibitem[{{Kawaler}(2015)}]{2015kawaler}
{Kawaler}, S.~D. 2015, in Astronomical Society of the Pacific Conference
  Series, Vol. 493, 19th European Workshop on White Dwarfs, ed. P.~{Dufour},
  P.~{Bergeron}, \& G.~{Fontaine}, 65

\bibitem[{{Kippenhahn}(1974)}]{Kippenhahn1974}
{Kippenhahn}, R. 1974, in IAU Symposium, Vol.~66, Late Stages of Stellar
  Evolution, ed. R.~J. {Tayler} \& J.~E. {Hesser}, 20

\bibitem[{Langer {et~al.}(1999)Langer, Heger, Wellstein, \&
  Herwig}]{Langer1999}
Langer, N., Heger, A., Wellstein, S., \& Herwig, F. 1999, \aap, 346, L37

\bibitem[{Lau {et~al.}(2012)Lau, Gil-Pons, Doherty, \& Lattanzio}]{Lau2012}
Lau, H. H.~B., Gil-Pons, P., Doherty, C., \& Lattanzio, J. 2012, \aap, 542, A1

\bibitem[{{Maeder}(1987)}]{1987Maeder}
{Maeder}, A. 1987, \aap, 178, 159

\bibitem[{Maeder \& Meynet(2000)}]{2000ARA&AMandM}
Maeder, A. \& Meynet, G. 2000, \araa, 38, 143

\bibitem[{Maeder \& Meynet(2012)}]{ReviewMandM}
Maeder, A. \& Meynet, G. 2012, Rev. Mod. Phys., 84, 25

\bibitem[{Marques {et~al.}(2013)Marques, Goupil, Lebreton, Talon, Palacios,
  Belkacem, Ouazzani, Mosser, Moya, Morel, Pichon, Mathis, Zahn,
  Turck-Chi{\`{e}}ze, \& Nghiem}]{Marques2013}
Marques, J.~P., Goupil, M.~J., Lebreton, Y., {et~al.} 2013, \aap, 549, A74

\bibitem[{{Mosser} {et~al.}(2012){Mosser}, {Goupil}, {Belkacem}, {Marques},
  {Beck}, {Bloemen}, {De Ridder}, {Barban}, {Deheuvels}, {Elsworth}, {Hekker},
  {Kallinger}, {Ouazzani}, {Pinsonneault}, {Samadi}, {Stello}, {Garc{\'{\i}}a},
  {Klaus}, {Li}, {Mathur}, \& {Morris}}]{2012Mosser}
{Mosser}, B., {Goupil}, M.~J., {Belkacem}, K., {et~al.} 2012, \aap, 548, A10

\bibitem[{{Nucci} \& {Busso}(2014)}]{2014Nucci}
{Nucci}, M.~C. \& {Busso}, M. 2014, \apj, 787, 141

\bibitem[{{Paxton} {et~al.}(2011){Paxton}, {Bildsten}, {Dotter}, {Herwig},
  {Lesaffre}, \& {Timmes}}]{2011_MESA_1}
{Paxton}, B., {Bildsten}, L., {Dotter}, A., {et~al.} 2011, \apjs, 192, 3

\bibitem[{{Paxton} {et~al.}(2013){Paxton}, {Cantiello}, {Arras}, {Bildsten},
  {Brown}, {Dotter}, {Mankovich}, {Montgomery}, {Stello}, {Timmes}, \&
  {Townsend}}]{2013_MESA_2}
{Paxton}, B., {Cantiello}, M., {Arras}, P., {et~al.} 2013, \apjs, 208, 4

\bibitem[{{Piersanti} {et~al.}(2013){Piersanti}, {Cristallo}, \&
  {Straniero}}]{2013_fruity_rotation}
{Piersanti}, L., {Cristallo}, S., \& {Straniero}, O. 2013, \apj, 774, 98

\bibitem[{{Pignatari} {et~al.}(2016){Pignatari}, {Herwig}, {Hirschi},
  {Bennett}, {Rockefeller}, {Fryer}, {Timmes}, {Ritter}, {Heger}, {Jones},
  {Battino}, {Dotter}, {Trappitsch}, {Diehl}, {Frischknecht}, {Hungerford},
  {Magkotsios}, {Travaglio}, \& {Young}}]{2016Pignatari}
{Pignatari}, M., {Herwig}, F., {Hirschi}, R., {et~al.} 2016, \apjs, 225, 24

\bibitem[{{Pinsonneault} {et~al.}(1989){Pinsonneault}, {Kawaler}, {Sofia}, \&
  {Demarque}}]{pin89}
{Pinsonneault}, M.~H., {Kawaler}, S.~D., {Sofia}, S., \& {Demarque}, P. 1989,
  \apj, 338, 424

\bibitem[{{Prantzos} {et~al.}(2018){Prantzos}, {Abia}, {Limongi}, {Chieffi}, \&
  {Cristallo}}]{Prantzos2018}
{Prantzos}, N., {Abia}, C., {Limongi}, M., {Chieffi}, A., \& {Cristallo}, S.
  2018, \mnras, 476, 3432

\bibitem[{{Siess} {et~al.}(2004){Siess}, {Goriely}, \& {Langer}}]{2004Siess}
{Siess}, L., {Goriely}, S., \& {Langer}, N. 2004, \aap, 415, 1089

\bibitem[{{Spada} {et~al.}(2016){Spada}, {Gellert}, {Arlt}, \&
  {Deheuvels}}]{2016Spada}
{Spada}, F., {Gellert}, M., {Arlt}, R., \& {Deheuvels}, S. 2016, \aap, 589, A23

\bibitem[{{Straniero} {et~al.}(1995){Straniero}, {Gallino}, {Busso}, {Chiefei},
  {Raiteri}, {Limongi}, \& {Salaris}}]{1995Straniero}
{Straniero}, O., {Gallino}, R., {Busso}, M., {et~al.} 1995, \apjl, 440, L85

\bibitem[{{Straniero} {et~al.}(2006){Straniero}, {Gallino}, \&
  {Cristallo}}]{2006_Straniero_Gallino_Cristallo}
{Straniero}, O., {Gallino}, R., \& {Cristallo}, S. 2006, Nuclear Physics A,
  777, 311

\bibitem[{{Suijs} {et~al.}(2008){Suijs}, {Langer}, {Poelarends}, {Yoon},
  {Heger}, \& {Herwig}}]{2008Suijs}
{Suijs}, M.~P.~L., {Langer}, N., {Poelarends}, A.-J., {et~al.} 2008, \aap, 481,
  L87

\bibitem[{Sweigart(1999)}]{Sweigart1999}
Sweigart, A.~V. 1999, in IAU Symposium, Vol. 190, New Views of the Magellanic
  Clouds, ed. Y.-H. {Chu}, N.~{Suntzeff}, J.~{Hesser}, \& D.~{Bohlender}, 370

\bibitem[{{Tayar} \& {Pinsonneault}(2013)}]{2013tayar}
{Tayar}, J. \& {Pinsonneault}, M.~H. 2013, \apjl, 775, L1

\bibitem[{{Tayar} \& {Pinsonneault}(2018)}]{2018tayar}
{Tayar}, J. \& {Pinsonneault}, M.~H. 2018, \apj, 868, 150

\bibitem[{{Travaglio} {et~al.}(2004){Travaglio}, {Gallino}, {Arnone}, {Cowan},
  {Jordan}, \& {Sneden}}]{Travaglio2004}
{Travaglio}, C., {Gallino}, R., {Arnone}, E., {et~al.} 2004, \apj, 601, 864

\bibitem[{von Zeipel(1924)}]{Zeipel1924}
von Zeipel, H. 1924, \mnras, 84, 665

\bibitem[{{Wallner} {et~al.}(2016){Wallner}, {Bichler}, {Buczak}, {Dillmann},
  {K{\"a}ppeler}, {Karakas}, {Lederer}, {Lugaro}, {Mair}, {Mengoni},
  {Sch{\"a}tzel}, {Steier}, \& {Trautvetter}}]{2016Wallner}
{Wallner}, A., {Bichler}, M., {Buczak}, K., {et~al.} 2016, \prc, 93, 045803

\bibitem[{{Wasiutynski}(1946)}]{Wasiutynski1946}
{Wasiutynski}, J. 1946, Astrophysica Norvegica, 4, 1

\bibitem[{Wood \& Faulkner(1986)}]{Wood1986}
Wood, P.~R. \& Faulkner, D.~J. 1986, \apj, 307, 659

\bibitem[{{Yoon} {et~al.}(2006){Yoon}, {Langer}, \& {Norman}}]{2006Yoon}
{Yoon}, S.~C., {Langer}, N., \& {Norman}, C. 2006, \aap, 460, 199

\bibitem[{{Zahn}(1974)}]{Zahn1974}
{Zahn}, J.-P. 1974, in IAU Symposium, Vol.~59, Stellar Instability and
  Evolution, ed. P.~{Ledoux}, A.~{Noels}, \& A.~W. {Rodgers}, 185--194

\end{thebibliography}

\begin{appendix}
\section{The s-process in models without additional viscosity}
\label{sec:app_sproc_noNUadd}

\begin{figure*}[h]%
    \centering 
    \subfloat{{\includegraphics[width=4cm]{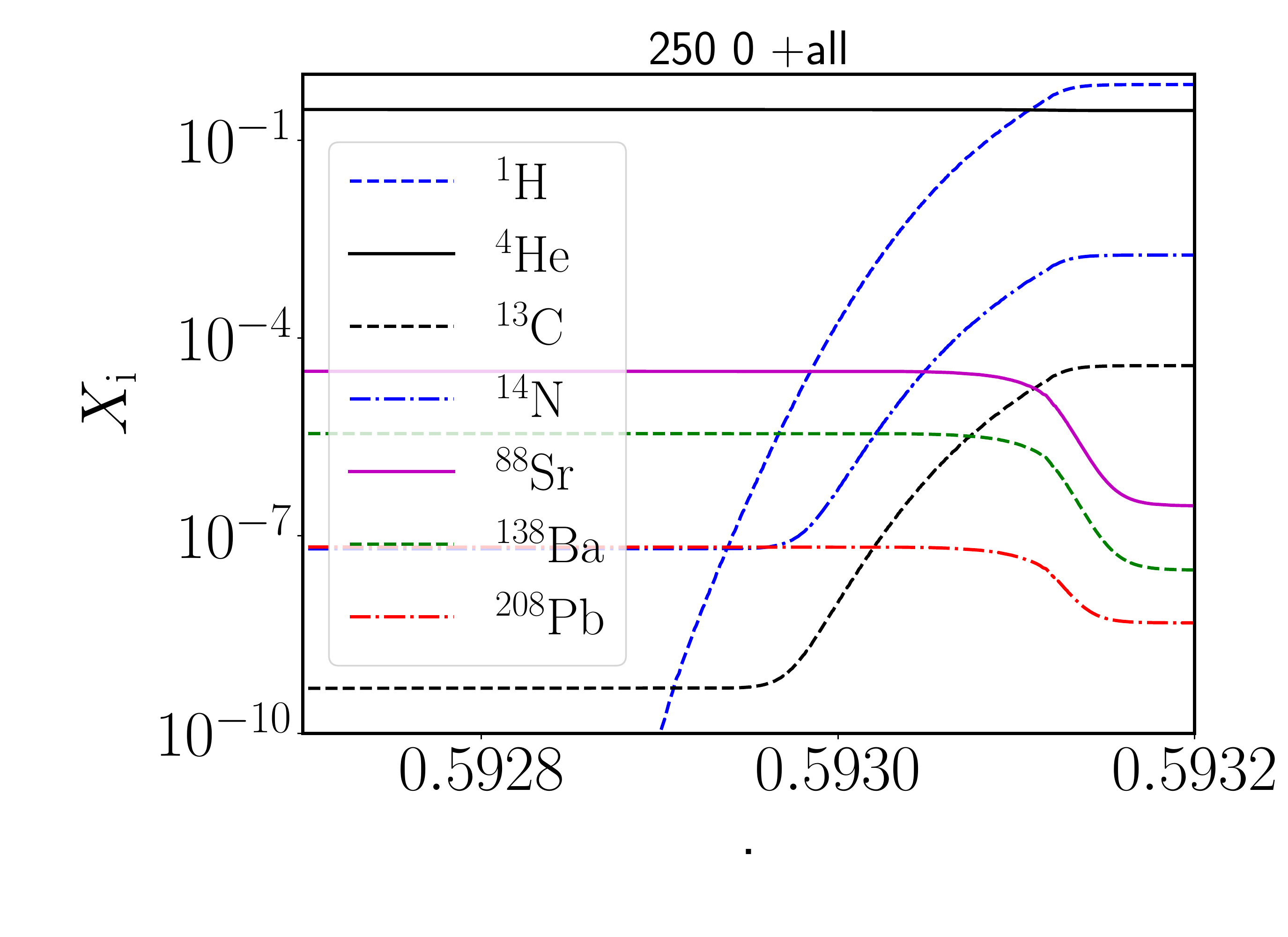}}}%
    \subfloat{{\includegraphics[width=4.5cm]{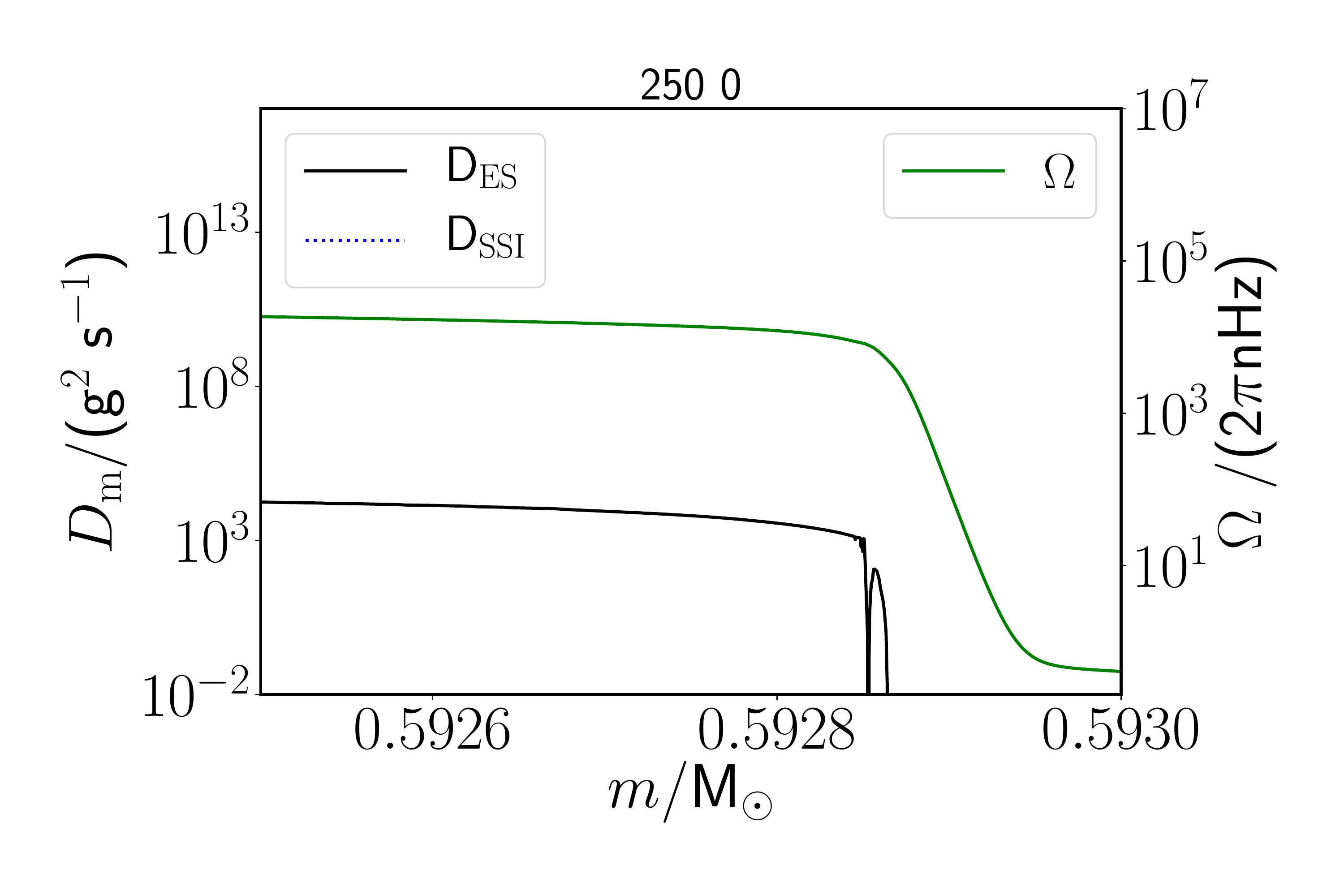}}}%
    \subfloat{{\includegraphics[width=4.5cm]{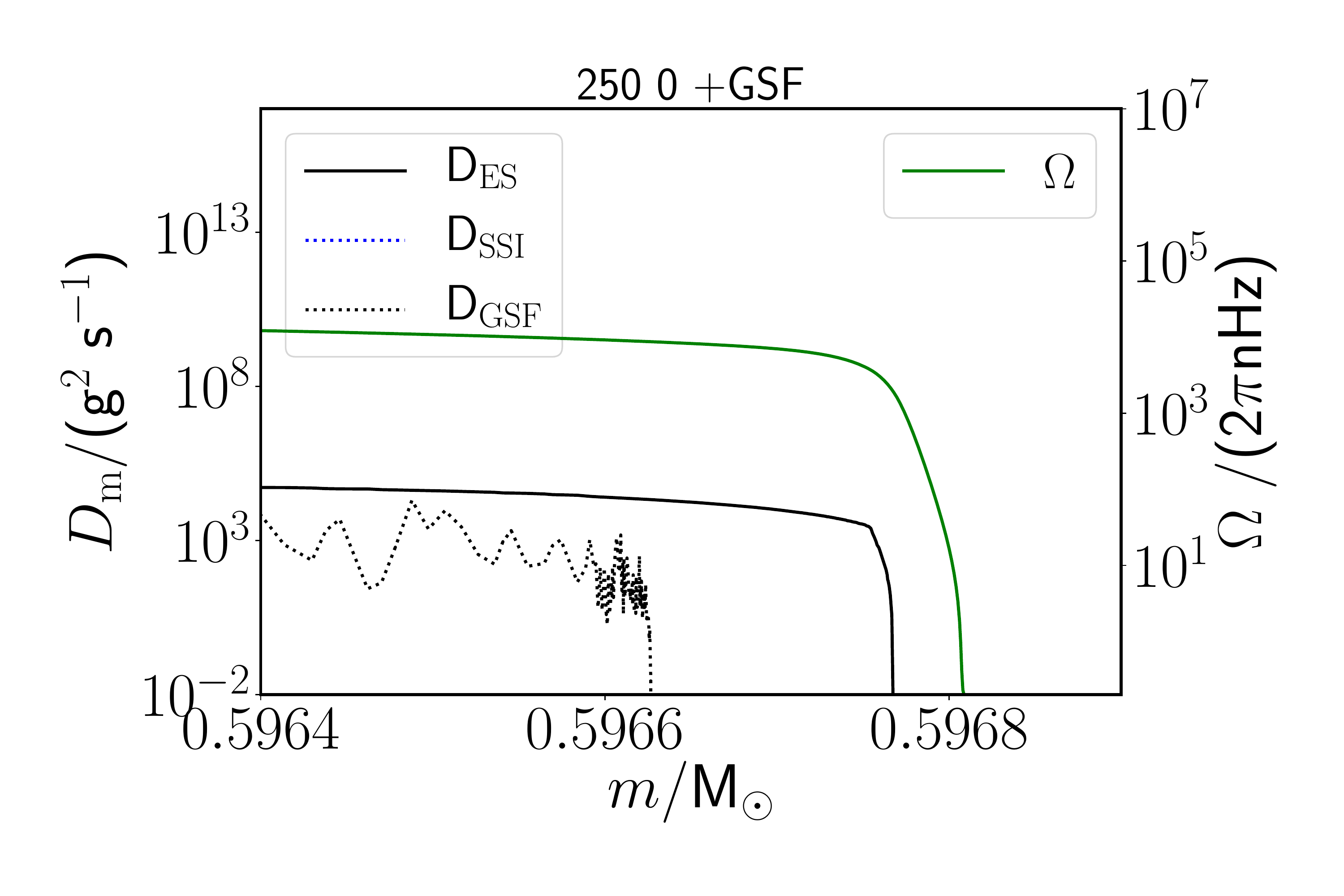}}}%
    \subfloat{{\includegraphics[width=4.5cm]{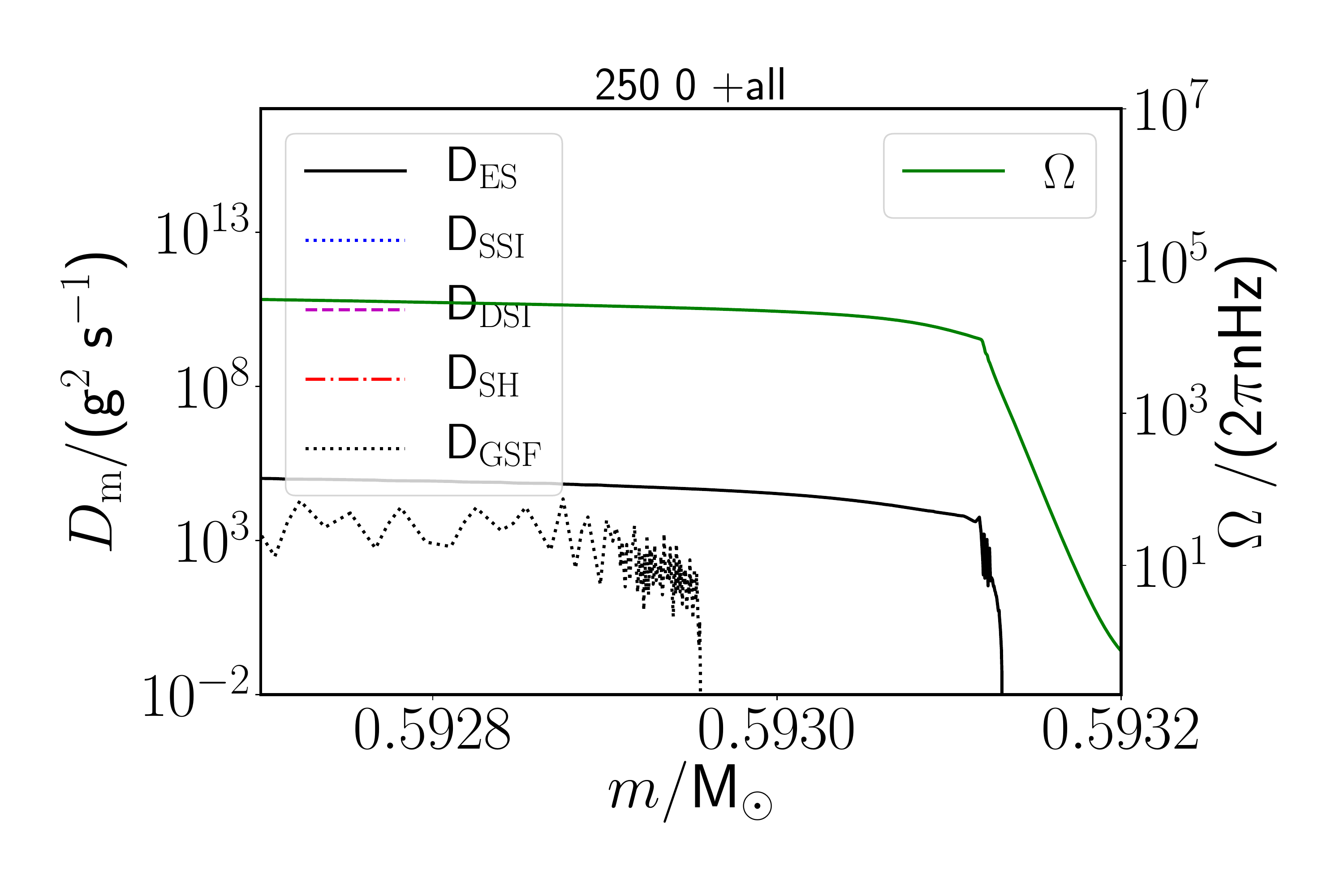}}}\\
    \vspace{-0.88cm}
    \subfloat{{\includegraphics[width=4cm]{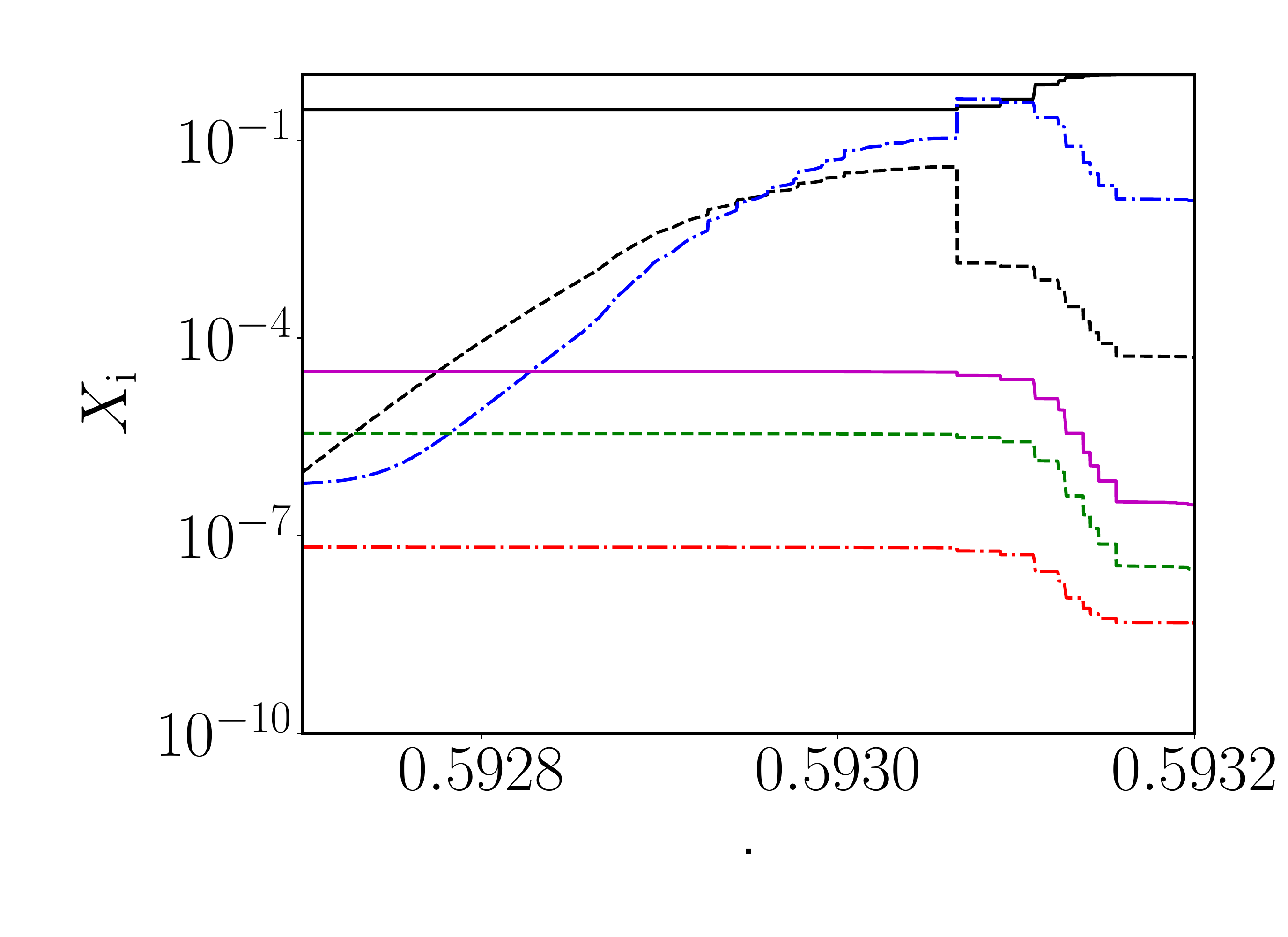}}}%
    \subfloat{{\includegraphics[width=4.5cm]{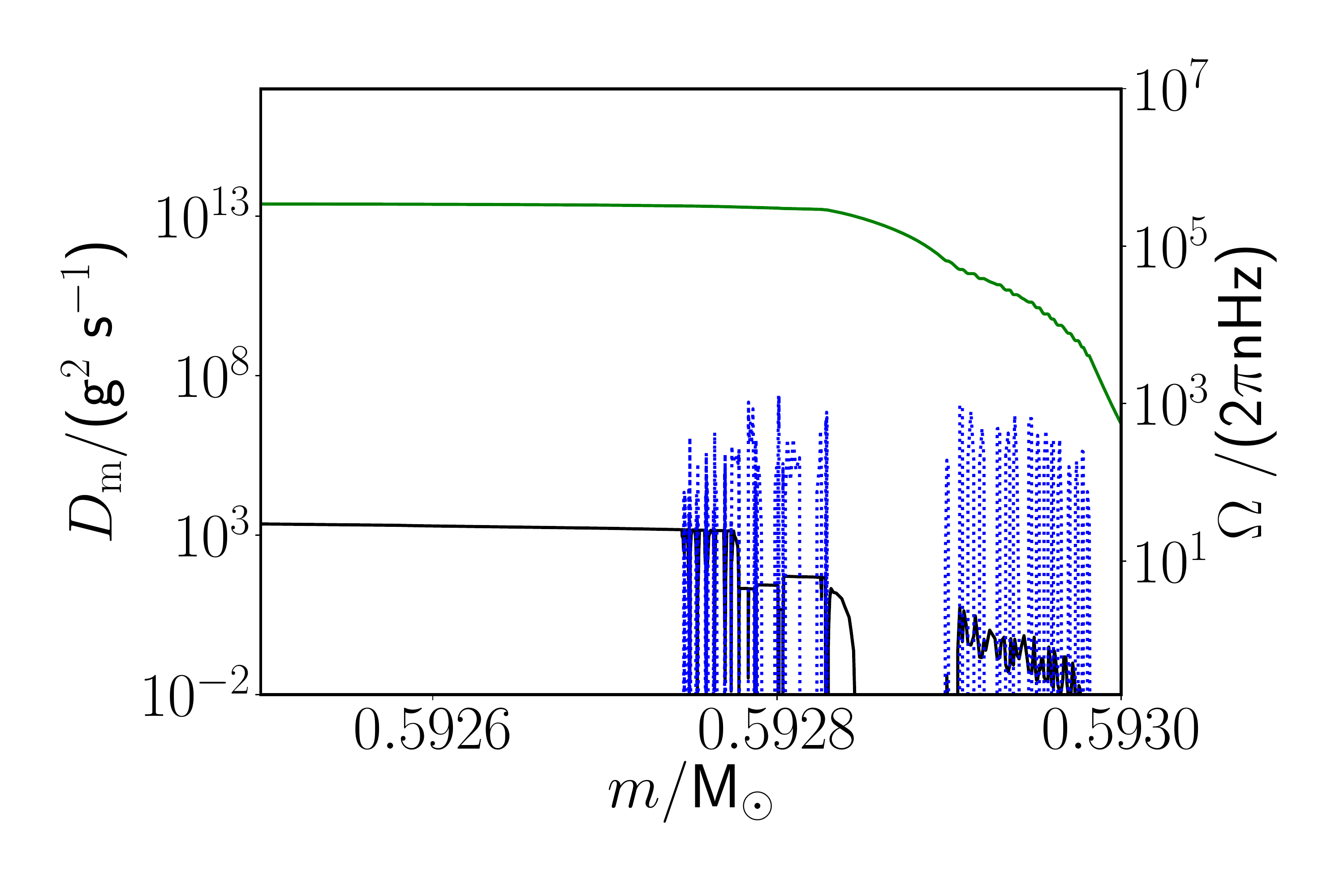}}}%
    \subfloat{{\includegraphics[width=4.5cm]{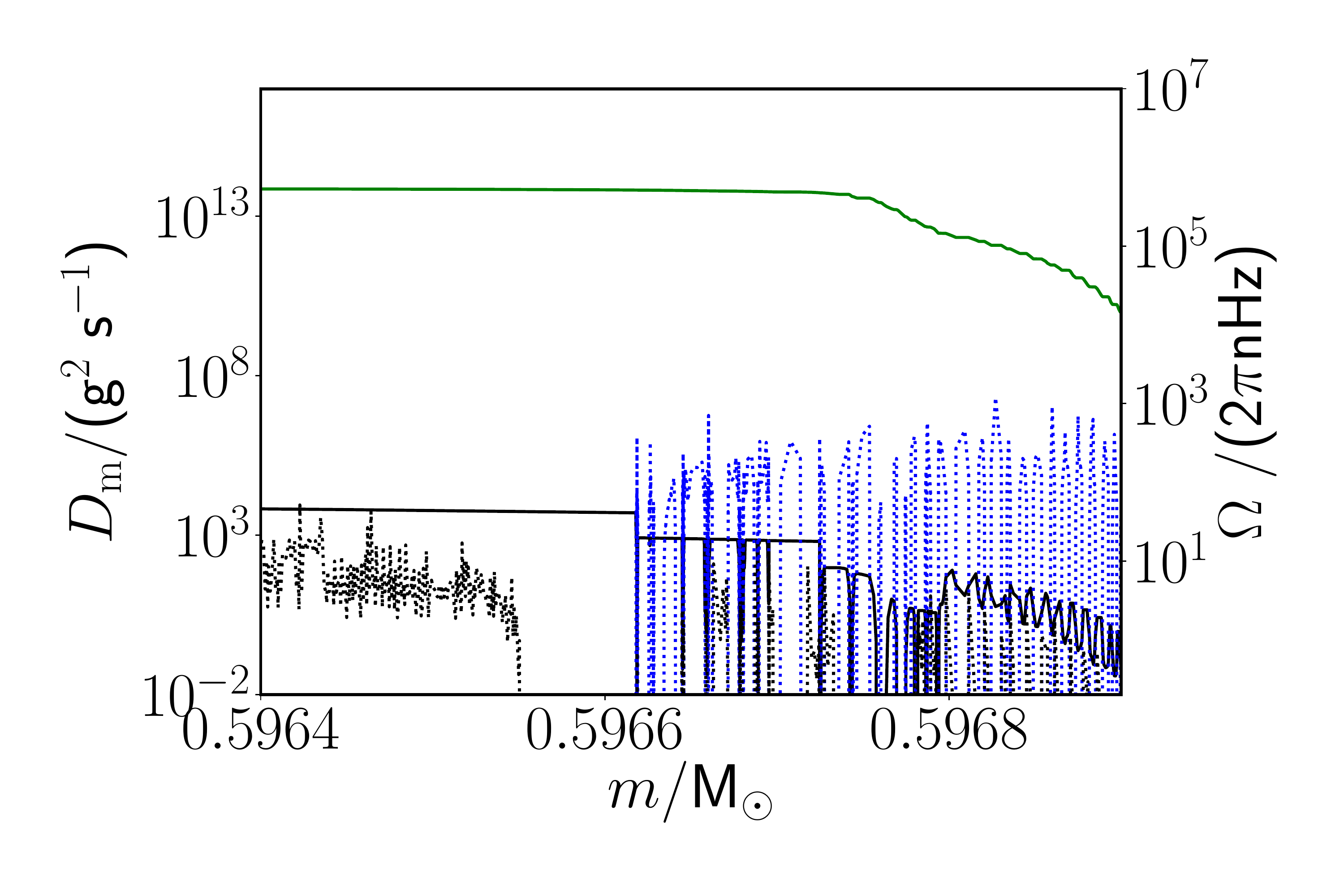}}}%
    \subfloat{{\includegraphics[width=4.5cm]{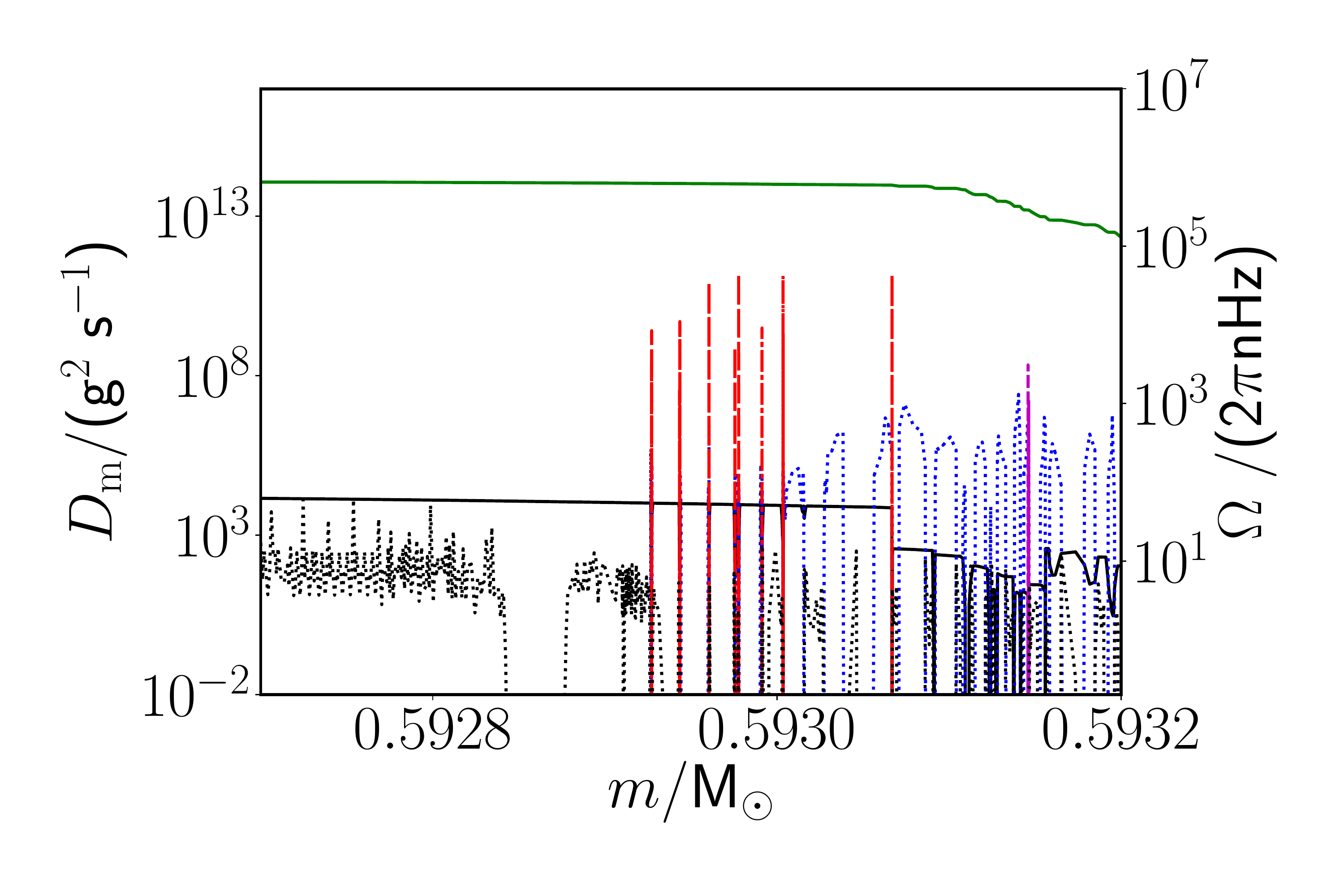}}}\\
    \vspace{-0.88cm}
    \subfloat{{\includegraphics[width=4cm]{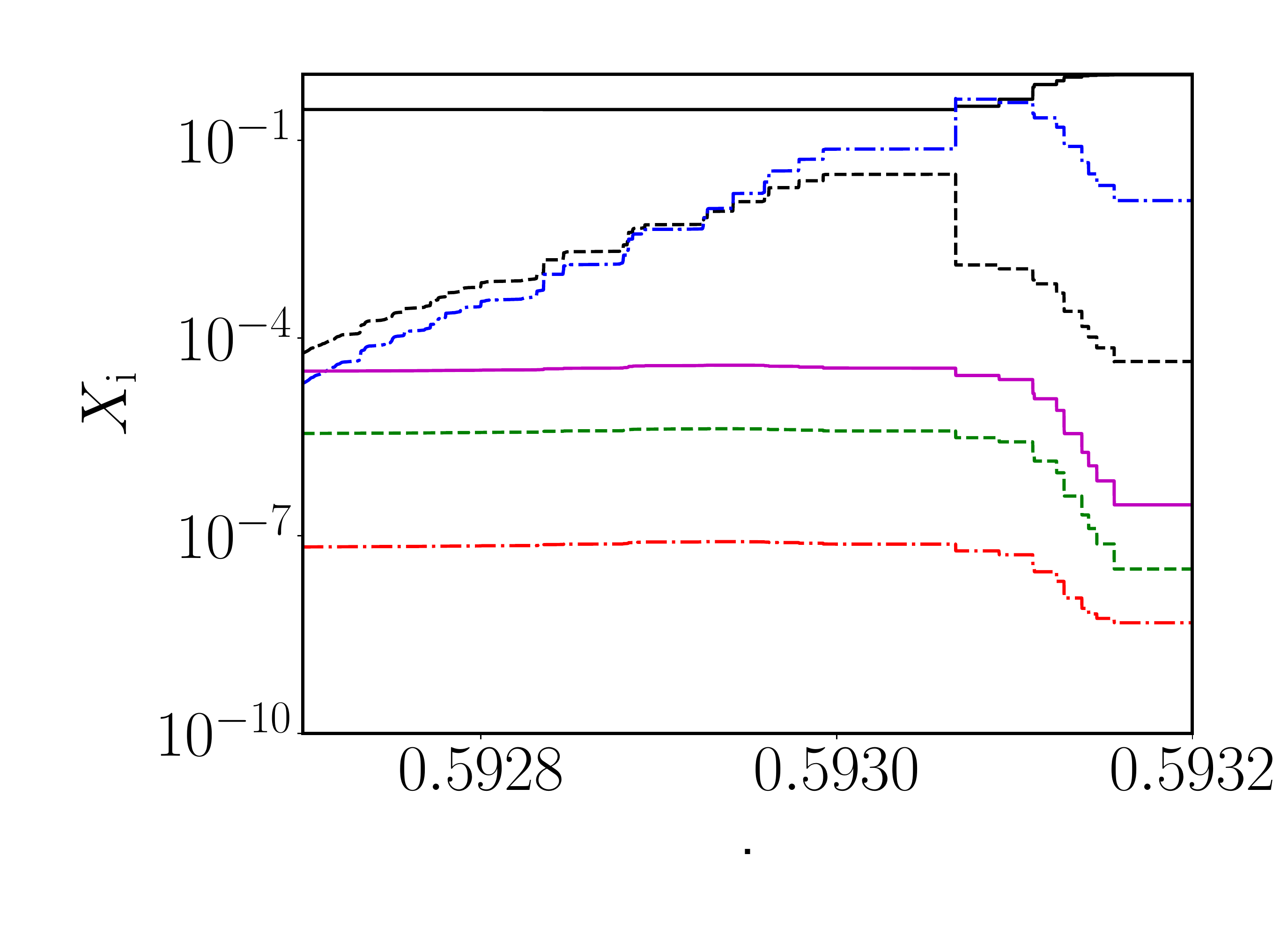}}}%
    \subfloat{{\includegraphics[width=4.5cm]{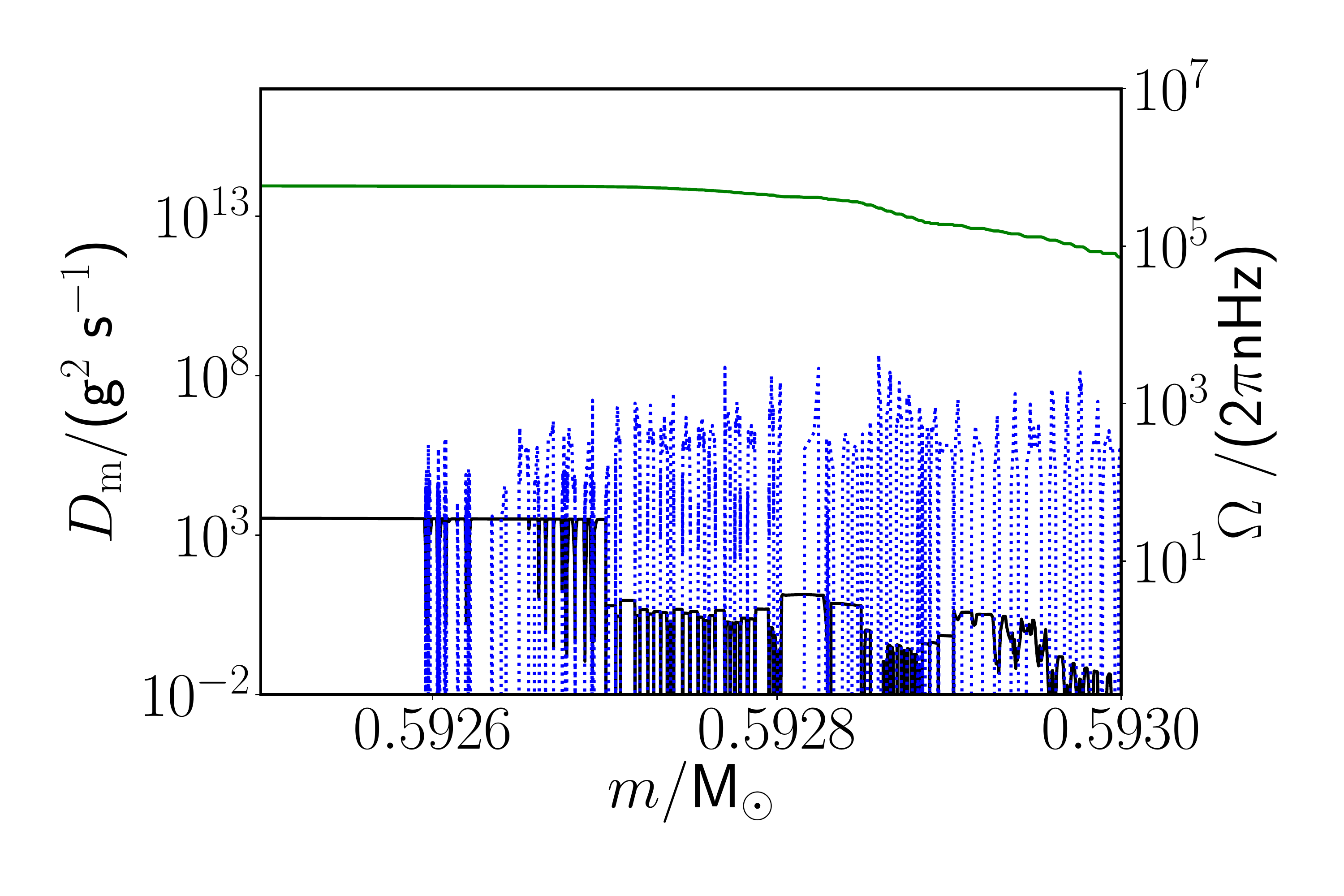}}}%
    \subfloat{{\includegraphics[width=4.5cm]{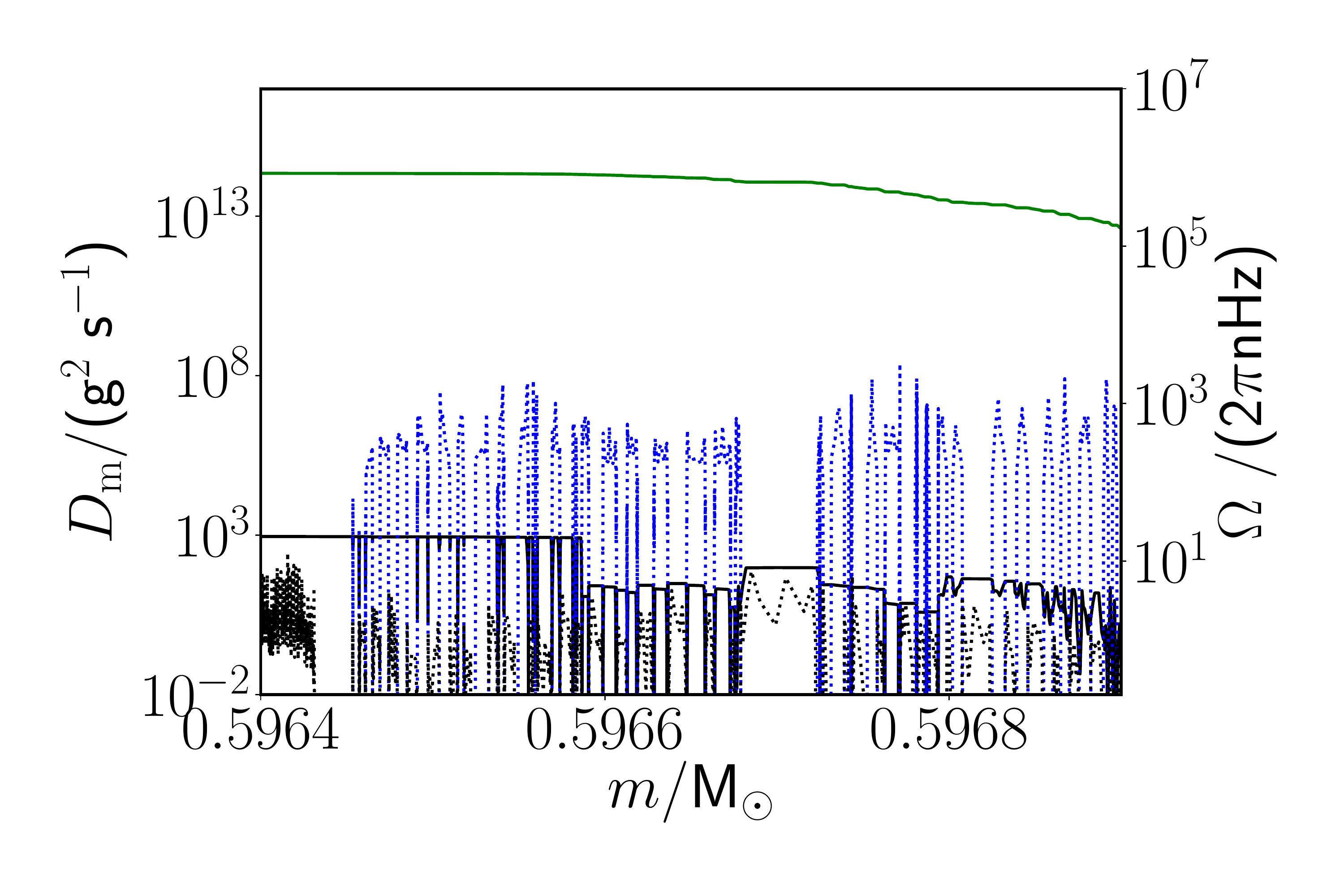}}}%
    \subfloat{{\includegraphics[width=4.5cm]{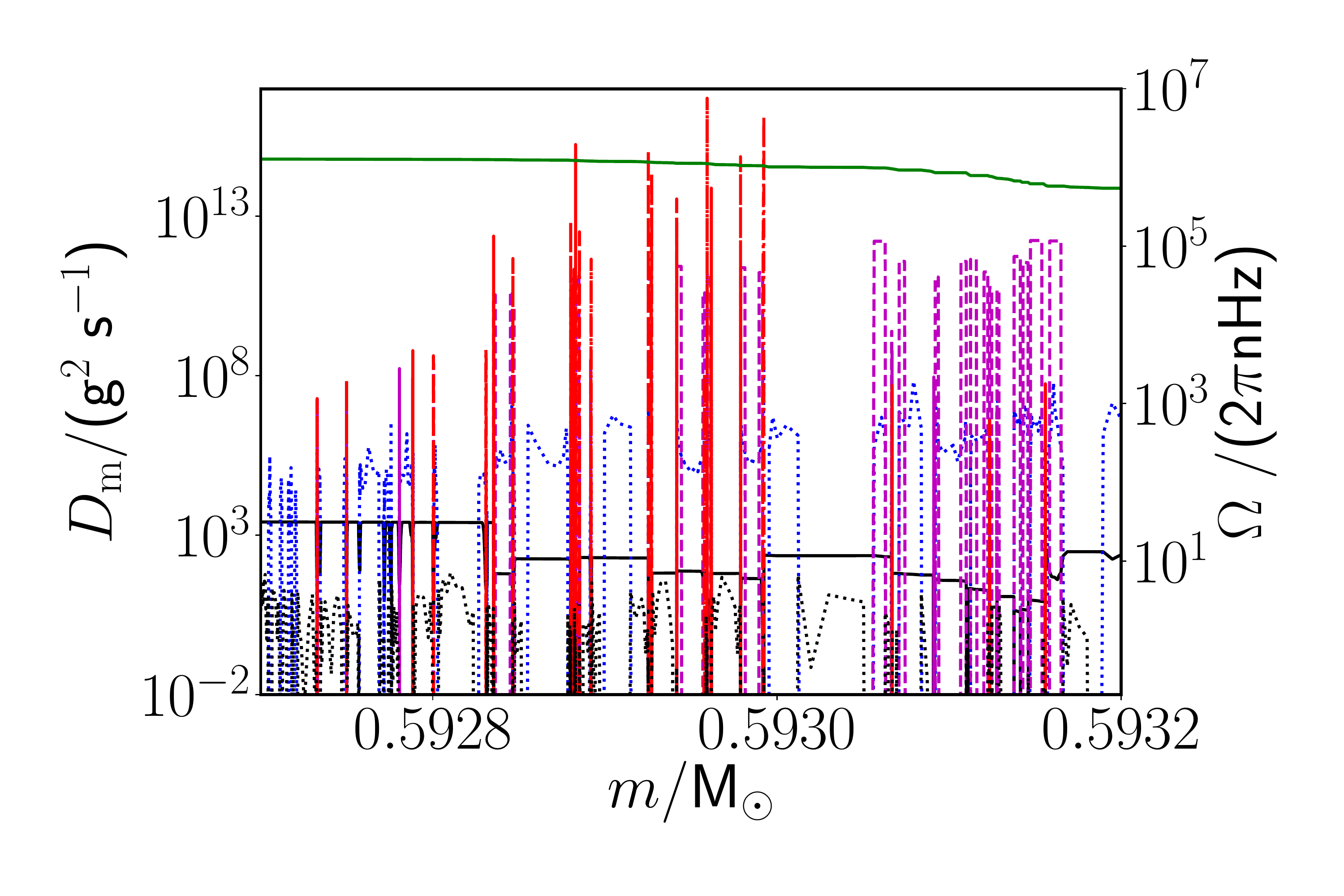}}}%
    \caption{Abundance and diffusion plots showing the maximum extent of the TDU (top panels), the maximum $^{13}$C-pocket size (middle panels), and the start of the the s-process production (bottom panels), matching the interpulse in which the fifth TDU takes place as in Fig.\thinspace\ref{fig:c13pckt_noR}. We show the abundance profiles of a characteristic $^{13}$C-pocket in the most left panel, followed by the diffusion profiles of the `250 0', `250 0 +GSF', and the `250 0 +all' models respectively. The $^{13}$C-pockets in these models are widened when compared to the `250 5' model in Fig.\thinspace\ref{fig:c13pckt_noR} due to the ES circulation. Please note the changed range over which the diffusion coefficients are shown compared to Fig.\thinspace\ref{fig:c13pckt_noR}.}
    \label{fig:c13_fastrot}%
\end{figure*}

In this appendix we describe our models that rotate too fast to match asteroseismically measured core rotation rates and provide a comparison to the previously published papers \citep{2003ApJHerwig_rot,2004Siess,2013_fruity_rotation}. We stress that for all these models the core rotate orders of magnitude faster in the evolved evolutionary phases, as compared to the observations. \\
Another difference between the models described in the main text and those presented here is the amount of rotationally induced mixing processes. Because in the previously published papers mentioned above all rotationally induced mixing processes as defined by \citet{2000ApJHeger} were included, we provide here a model that also includes all processes. \citet{2013_fruity_rotation} mentions that the GSF instability is the main process responsible for the pollution of the $^{13}$C-pocket by $^{14}$N, limiting the neutron exposure and keeping the s-process production concentrated around the Sr/Y/Zr peak. We therefore also add a model that includes only the ES circulation, the SSI, and the GSF instability. The three models described in this Appendix is listed in Table \ref{tab:app_B}.

\begin{table}
    \centering
    \caption{Set of models described in this Appendix. Only the rotational instabilities are listed as all other parameters are equal.}
    \begin{tabular}{l|ccccr}
     Model    &  ES & SSI & GSF & DSI & SH\\
     \hline
     250 0      & y & y & - & - & -\\
     250 0 +GSF & y & y & y & - & -\\
     250 0 +all & y & y & y & y & y
    \end{tabular}

    \label{tab:app_B}
\end{table}

\subsection{Effects on the $^{13}$C-pocket of the inclusion of all rotationally induced diffusion processes}

The two new models are restarted from the `250 0' model in Table \ref{tab:mass} at the last TP before the first TDU and thus before the first $^{13}$C-pocket. This allows for a direct comparison of s-process production in these models to the `250 0' model without the extra mixing processes, as the first TDU is the start of the s-process production.\\
The abundance profiles shown in the left column of Fig.\thinspace\ref{fig:c13_fastrot} are characteristic for the models presented in this section. Compared to the abundance profiles of the `250 5' model, there are two distinct differences. The first is that the $^{13}$C-pocket in Fig.\thinspace\ref{fig:c13_fastrot} is widened compared to the $^{13}$C-pocket in Fig.\thinspace\ref{fig:c13pckt_noR} . This is due to the higher rotation rate leading to the ES circulation being two orders of magnitude stronger in the `250 0' models, see columns 2$-$4 in Fig.\thinspace\ref{fig:c13_fastrot}, than in the `250 5' model. The second difference is that the abundance profiles in the `250 0' pocket are less smooth than in the `250 5' pocket. This is due to the discontinuous mixing by the SSI, as already mentioned in Sect.\thinspace\ref{sec:noR_pckt}. The ES circulation is however still present in the $^{13}$C-pocket region in the `250 0' model even when the s-process production has started. This results in poisoning of the `250 0' pocket by $^{14}$N.\\
The diffusion profiles of the model including the GSF instability are shown in the third column from the left in Fig.\thinspace\ref{fig:c13_fastrot}. This instability depends on both the $\Omega$ values and on the spatial derivative of $\Omega$, and is present almost throughout the mass range shown. It is however not dominant over the ES circulation or the SSI, and will therefore not have much effect on the s-process production, contrarily to what was concluded by \citet{2013_fruity_rotation}.\\
The right column in Fig.\thinspace\ref{fig:c13_fastrot} shows the diffusion profiles of the model including all rotationally induced instabilities. Both new instabilities (DSI and SH) have diffusion profiles with a discrete character and will therefore have limited effect on the s-process production within this model.

   \begin{figure}[ht]
   \centering
   \includegraphics[width=\linewidth]{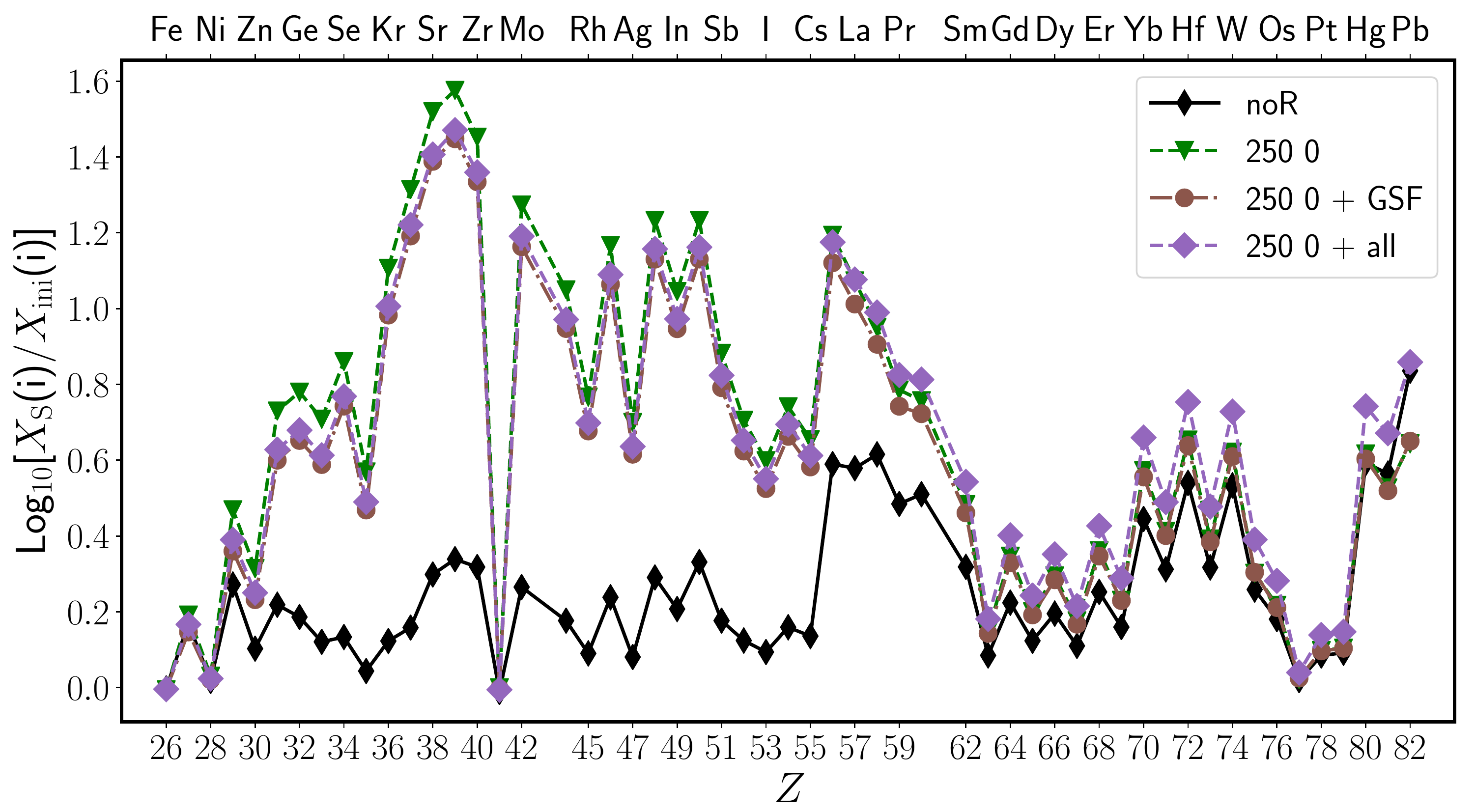}
   \caption{Surface enrichment of the non rotating and the 250 0 models, showing that the inclusion of GSF, SH and DSI does not alter the s-process production. This is a numerical issue: there is work to be done between the derivation of the instabilities and their implementation in stellar evolutionary codes.}
   \label{fig:surf_enrich_fast}%
   \end{figure}
   
\subsection{Surface enrichment}

Figure \ref{fig:surf_enrich_fast} shows the comparison of the surface enrichments, including the surface enrichment of the non-rotating model. All `250 0' models are comparable in this figure, confirming the findings of the previous section that the inclusion of GSF, DSI, and SH does not have an effect on the s-process production. Compared to the non-rotating model, the s-process production has greatly increased up to Sm. We thus also find that rotation could increase the s-process production. This increase can be explained by the widened $^{13}$C-pocket, allowing for more Fe-group seeds to be activated by neutron captures. The pocket is widened compared to the non-rotating models because of the ES circulation being active during the creation of the pocket. The poisoning of the $^{13}$C-pocket by ES circulation mixing in $^{14}$N is the reason why this increased production has not continued until Pb. \\
The surface enrichment of the models included in Fig.\thinspace\ref{fig:surf_enrich_fast} can be compared to \citet{2013_fruity_rotation} as they present 2 M$_{\odot}$ models at solar metallicity albeit at much slower rotation rates. The trends these models show is that the inclusion of rotation reduces the overall s-process production, due to the contamination of the pocket by $^{14}$N, which is opposite to what we find and further investigation would be needed to understand this difference. However, both sets of rotating models show core rotation rates that are several order of magnitude above the asteroseismically measured rotation rates throughout the evolution. Further studies do not seem warranted.\\\
Comparison to \citet{2003ApJHerwig_rot} and \citet{2004Siess} is less straightforward, as the first study concludes that the combination of overshoot (now renamed as convective boundary mixing) and rotation might allow for a spread in s-process production in AGB stars, while the second study does not combine the two processes.\\
Neither of the previously published studies on rotating AGB stars mentioned the changes in smoothness of abundance profiles as reported in the previous section.

\subsection{Discontinuous mixing and smoothing options}
The reason why we find these differences with \citet{2013_fruity_rotation},
may be related to the choice of smoothing options. The discontinuous character of several instabilities are caused by two features within the implementation of the instabilities. The first is that the implementation itself of these instabilities allows for a discontinuous behaviour, as there is of course a stability criteria present in the implementation. If the zone within a model is unstable according to the instability criteria, the instability becomes active, while in the next zone it can be stable again. The second issue is that when dynamical and secular shear appear, they should be taken into consideration immediately and not at the start of the next time step. The current implementation does include the shear at the next time step and therefore overestimates its impact. These issues reduce the practical use of these instabilities \citep[as also concluded by][in a different astrophysical context]{Aerts2018a}.\\
Smoothing options are available and tested to solve the issues, however, it is impossible to decide which feature is physical and should not be smoothed, and which is numerical and should be smoothed. Therefore, in this work we have decided to avoid the use of smoothing functions. Among several different options tested, the only `smoothing' option that seems to  effectively improve stellar profiles is the inclusion of a low additional viscosity. This has the effect that the $\Omega$-profile is smoothed, which leads to a reduced appearance of secular and dynamical shear. However, the discontinuous behaviour of the SH instability is still present. Including all instabilities in an accurate manner remains a challenge.

\end{appendix}

\end{document}